\documentclass[12pt]{iopart} 
\usepackage{color} 
\usepackage{graphics} 
\usepackage{amssymb} 
\def\neq {\not\equiv}

\def\cs2{c_{s}^{2}}

\def\p{\partial}

\def\be   {\begin{equation}}   \def\ee   {\end{equation}} 
\def\ba   {\begin{array}}      \def\ea   {\end{array}} 
\def\bea  {\begin{eqnarray}}   \def\eea  {\end{eqnarray}} 
\def\bean {\begin{eqnarray*}}  \def\eean {\end{eqnarray*}} 
\begin{document}

\title{An Estimator for statistical anisotropy from the CMB bispectrum} 

\vspace{0.8cm} 

\author{N.~Bartolo$^{1,2}$, E.~Dimastrogiovanni$^{1,2}$, M.~Liguori$^{4}$, S.~Matarrese$^{1,2}$ and A.~Riotto$^{2,5}$} 
\vspace{0.8cm} 
\address{$^1$ Dipartimento di Fisica ``G. Galilei'', Universit\`{a} degli Studi di 
Padova, \\ via Marzolo 8, I-35131 Padova, Italy} 
\address{$^2$ INFN, Sezione di Padova, via Marzolo 8, I-35131 Padova, Italy} 
\address{$^4$ Institut d'Astrophysique de Paris, 
UMR-7095 du CNRS, Universit´e Pierre et Marie Curie, 
98 bis bd Arago, 75014 Paris, France} 
\address{$^5$ CERN, Theory Division, CH-1211 Geneva 23, Switzerland} 
\vspace{0.4cm} 

\date{\today} 
\vspace{1cm} 

\begin{abstract} 


\noindent 
Various data analyses of the Cosmic Microwave Background (CMB) provide observational hints of statistical isotropy breaking. 
Some of these features can be studied within the framework of 
primordial vector fields in inflationary theories which generally display some level of statistical anisotropy both in the power spectrum and in higher-order correlation functions.   
Motivated by these observations and the recent theoretical developments in the study of primordial vector fields, 
we develop the formalism necessary to extract statistical anisotropy information from the three-point function of the CMB temperature anisotropy.     
We employ a simplified vector field model and parametrize the bispectrum of curvature fluctuations in such a way that all the information about 
statistical anisotropy is encoded in some parameters $\lambda_{LM}$ (which measure the anisotropic to the isotropic bispectrum amplitudes). For such a template bispectrum, we compute an optimal estimator for $\lambda_{LM}$ and the expected signal-to-noise ratio. 
We estimate that, for $f_{NL}\simeq 30$, an experiment like Planck can be sensitive to a ratio of the anisotropic to the isotropic amplitudes of the bispectrum as small as $10\%$. Our results are complementary to the information coming from a power spectrum analysis and 
particularly relevant for those models where statistical anisotropy turns out to be suppressed in the power spectrum but not negligible in the bispectrum.   

\end{abstract} 

\maketitle 

\section{Introduction} 

Cosmic Microwave Background (CMB) anisotropies have so far been analysed under the assumption of a statistically isotropic distribution. However, recent observations of ``anomalous'' features in the temperature WMAP maps might be interpreted as an indication of statistical anisotropy (although the a posteriori choice of statistics could make their interpretation difficult~\cite{Bennett2011}). These features concern (see \cite{Copietal} for a recent review): a ``cold spot'' in the southern galactic hemisphere \cite{Vielva:2003et,Cruz:2006fy}; an alignment between the quadrupole and octupole modes in the temperature anisotropies \cite{Bennett:1996ce,Spergel:2003cb,de OliveiraCosta:2003pu,Efstathiou:2003tv,Land:2005ad}; 
an asymmetry in the large scale power spectrum and in higher-order correlation functions between the northern and the southern ecliptic hemisphere \cite{Hansen:2004mj,Eriksen:2003db,Hansen:2004vq,Pietrobon:2009qg}. Other interesting analyses look for a quadrupolar modulation of the CMB power spectrum \cite{Groeneboom:2009,Hanson:2009gu,Groeneboom:2009cb}, which we discuss later in more details, see Eq.~(\ref{ps1})~\footnote{Let us also recall the lack of power of the temperature two-point correlation function 
on large angular scales ($> 60^\circ$), asymmetries in the even vs. odd multipoles of the CMB power spectra (parity symmetry breaking), both at large 
\cite{Kim:2010a,Gruppuso:2010} and intermediate angular scales (see for a discussion ~\cite{Bennett2011}, \cite{Pontzen:2010yw} and  Refs. therein). See also~\cite{Naselsky:2011ec}.

}. These experimental findings have been partially responsible for a renewed interest in models of the early Universe predicting some level of statistical anisotropy. 

Inflationary models where 
primordial vector fields play a non-negligible role are an interesting possibility. Models of this type have been proposed since 1989 \cite{Ford:1989me} in an attempt to achieve an exponential expansion for the early Universe without having to resort to scalar fields. A similar motivation inspired more recent papers \cite{Golovnev:2008cf,Golovnev:2008hv,Golovnev:2009ks,Maleknejad:2011jw,Maleknejad:2011sq}. Another possibility is provided by more general scenarios where vector fields do not necessarily drive inflation~\cite{Dimopoulos:2006ms,Ackerman:2007nb,Yokoyama:2008xw,Watanabe:2009ct, 
Dimopoulos:2008yv,Himmetoglu:2008hx,Himmetoglu:2009qi,Gumrukcuoglu:2010yc,Watanabe:2010fh,Kanno:2010nr,MurataSoda}. Some interest has also been directed to models of anisotropic dark energy, e.g. 
\cite{ArmendarizPicon:2004pm,Boehmer:2007qa,Koivisto:2007bp,Koivisto:2008xf,Jimenez:2009py,Jimenez:2009zza}. See also \cite{Graham:2010hh,BlancoPillado:2010uw} for other anisotropic scenarios. 

A general prediction of inflationary models where one (or more) vector field is present is that the power spectrum of primordial curvature perturbations acquires a directional dependence on the Fourier wavevector.     
A widely used parametrization is the following~\cite{Ackerman:2007nb} 
\footnote{ 
For a more general parametrization see~\cite{Pullen:2007tu}.} 
\bea\label{ps1} 
P(\vec{k})=P^{iso}(k)\left[1+G(k)\left(\hat{k}\cdot\hat{N}\right)^2\right], 
\eea     
where $G(k)$ is the (generally scale-dependent) amplitude of the statistical isotropy breaking induced by a preferred spatial direction $\hat{N}$. Notice that the isotropic part of the expression has been factorized in the function $P^{iso}(k)$. Such a parametrization of the power spectrum is valid for most vector field inflationary models. A minimum-variance estimator for the parameter $G$ was first built in \cite{Pullen:2007tu} showing that an experiment like Planck should be able to detect a quadrupolar anisotropy in the power spectrum as small as $0.5\%$ at 1$\sigma$ (see also~\cite{MA}). Afterwords, in \cite{Groeneboom:2009}, \cite{Hanson:2009gu} and \cite{Groeneboom:2009cb}, the five-year WMAP temperature data were analysed in order to estimate the magnitude and orientation of statistical anisotropy for a power spectrum with a form as in Eq.~(\ref{ps1}). Assuming a constant amplitude $G_*$, the authors of \cite{Groeneboom:2009cb} find $G_*=0.29\pm 0.031$. In \cite{Hanson:2009gu} and later in \cite{Hanson:2010gu}, the authors point out that the existence of beam asymmetries, uncorrected in the maps, should be accounted for and can be entirely responsible for the aforementioned power anisotropy (see however also the different conclusions of \cite{Groeneboom:2009cb}).
Anisotropic effects on CMB polarization and on the power spectrum of temperature fluctuations were also investigated in \cite{Dvorkin:2007jp,ArmendarizPicon:2005jh,Watanabe:2010bu,MA} and a study of a statistical anisotropic power spectrum for the temperature fluctuations has been presented in \cite{Aich:2010gn} accounting for anisotropies originated at the last scattering surface, in addition to the ones present in the primordial power spectrum. As far as constraints from Large-Scale-Structure analyses are concerned, Ref.~\cite{PH} considered a sample of photometric Luminous Red Galaxies from the SDSS survey to assess the quadrupole anisotropy in the primordial power spectrum of the type described by Eq.~(\ref{ps1}). Assuming the same preferred direction singled out by \cite{Groeneboom:2009}, they derive a constraint on the anisotropy amplitude $G_*=0.006 \pm 0.036$ ($1 \sigma$), thus finding no evidence for anisotropy.  Marginalizing over $\hat{N}$ with a uniform prior they find $-0.41 < G_* < 0.38$ at $95\%$ C.L. These results could confirm that the signal seen in CMB data is of systematic nature. However it must be stressed that CMB and LSS analyses probe different scales, and the amplitude of the anisotropy generally is scale dependent $G=G(k)$~\cite{Ackerman:2007nb}.

On the other hand, the study of inflationary models with vector fields has been also motivated by the possibility of predicting higher (compared to the standard slowly rolling scalar field model) levels of non-Gaussianity, combined with some level of statistical anisotropy in the three or higher-order correlation functions of primordial curvature perturbations (``Anisotropic non-Gaussianity'') \cite{Yokoyama:2008xw,ValenzuelaToledo:2009af,Dimopoulos:2008yv,ValenzuelaToledo:2009nq,Karciauskas:2008bc,Bartolo:2009pa,Bartolo:2009kg,Dimopoulos:2009am,Dimopoulos:2009vu} (see also \cite{Dimastrogiovanni:2010sm} for an overview of all of these models). 

In the spirit of \cite{Ackerman:2007nb,Pullen:2007tu,Hanson:2009gu} and motivated by a strong interest in anisotropic non-Gaussianities, we construct an estimator to extract 
statistical anisotropy information from the CMB bispectrum. We assume a primordial origin for violation of statistical isotropy as we investigated in \cite{Bartolo:2009pa,Bartolo:2009kg} 
where very general models of inflation with primordial vector fields have been considered including in particular possible contributions of non-Abelian $SU(2)$ gauge self-interactions. 
For the minimal and simplest version of vector field inflationary models, we study a specific bispectrum template for which the departure from statistical isotropy is quantified by some 
coefficients $\lambda_{LM}$. Essentially these parameters correspond to the ratio between the non-linearity amplitude of the anisotropic bispectrum and the isotropic one: $\lambda_{\rm LM} \sim ( f^{(A)}_{\rm NL}/f^{(I)}_{\rm NL})$ (see Appendix A). For such parameters we perform a Fisher matrix analysis to give an estimate of the potentiality of present and future experiments to put constraints on such anisotropic features. 

One of the motivations for our work is that 
an analysis of anisotropic non-Gaussianity can be complementary to that probing anisotropic features in the power spectrum, so as to offer a possible cross-check for statistical anisotropic signatures.  There is, however, another important motivation to point out. 
Various models for isotropy violation can indeed be characterized by a negligibly small (unobservable) level of statistical anisotropy in the power spectrum, but they still display a relevant anisotropic amplitude in the bispectrum (see, e.g.,~\cite{Yokoyama:2008xw,Bartolo:2009pa,Bartolo:2009kg}).  The essential reason is that the non-linearity parameter for the anisotropic bispectrum $f^{(A)}_{\rm NL}$ turns out generally to depend not solely on $G$ but also on other model-dependent parameters which can compensate for a parameter $|G| \ll 1$. 
For example the case of non-Abelian primordial vector fields \cite{Bartolo:2009pa,Bartolo:2009kg} is particularly clear since $f^{(A)}_{\rm NL}$ does not even depend on $G$ when the main contribution comes from the intrinsic self-interactions of the vector fields. This is because such self-interactions just determine the bispectrum without entering into the (tree-level) power spectrum. In other models $f^{(A)}_{\rm NL} \sim G^2  f({\bf A})$, where $f({\bf A})$ is some functions of the vector fields involved (one can check this is the case for the models discussed in, e.g., Refs.~\cite{Yokoyama:2008xw,Dimopoulos:2008yv}). Therefore, even when $|G| \ll 1$ is not observable through an analysis of the power spectrum, the parameters space of the models can still allow for a non-negligible non-linearity amplitude of the anisotropic bispectrum part. In fact generally it turns out that (even for unobservable level of $G$) $ |\lambda_{\rm LM}|  \sim |f^{(A)}_{\rm NL}/f^{(I)}_{\rm NL} | \leq 1$, i.e the anisotropic amplitude of the bispectrum can even be as large as the isotropic part (see Appendix A for details). 

In this paper we will quantify what is the level required for the CMB primordial bispectrum statistical anisotropy to be measured, showing that there might ample room for a high level of statistical anisotropy in the bispectrum when $|G| \ll 1$. Searches for signatures of statistical anisotropy limited to the power spectrum would thus be blind to such features.


The paper is organized as follows: in Sec.~2 we compute the three-point function of the temperature anisotropies harmonic coefficients, considering a simplified vector field model of inflation; in Sec.~3 we derive an optimal estimator for the anisotropy parameters; in Sec.~4 we estimate the signal-to-noise ratio; in Sec.~5 we draw our conclusions. In Appendix A we provide an overview of the bispectrum predictions
of different primordial vector field models and discuss the parametrization of the
bispectrum adopted in the paper, discussing also in some details a specific example;
Appendix B includes a brief discussion of Bayesian analysis for complex parameters; in Appendix C we discuss in details the order of the Edgeworth expansion that we have employed; finally,
Appendix D collects some useful properties of Wigner-3j and 6j symbols.

\section{Anisotropic CMB bispectrum} 

As discussed in the introduction, several inflationary models have been proposed that incorporate primordial vector fields. For many of these theories, the main outcome is the introduction of statistical anisotropy features in the correlation functions of curvature perturbations $\zeta$. We are interested in the three-point function, or bispectrum 
\bea\fl 
\langle\zeta(\vec{k}_{1})\zeta(\vec{k}_{2}) \zeta(\vec{k}_{3})\rangle=(2\pi)^3\delta^{(3)}(\vec{k}_{1}+\vec{k}_{2}+\vec{k}_{3})B(\vec{k}_{1},\vec{k}_{2},\vec{k}_{3})\, . 
\eea 
Similarly to the power spectrum, Eq.(\ref{ps1}), in these models also the bispectrum $B(\vec{k}_{1},\vec{k}_{2},\vec{k}_{3})$ exhibits a dependence on the angle between a preferred spatial direction(s) specified by some vector(s) and the orientation in space of the wave vectors $\vec{k}_{1}$, $\vec{k}_{2}$ and $\vec{k}_{3}$. For single vector field models a useful parametrization is (see Appendix A for a detailed overview of the models to which this parametrization applies) 
\bea\label{bis1} 
\fl 
B(\vec{k}_{1},\vec{k}_{2},\vec{k}_{3})&=&B^{iso}(k_{1},k_{2})\Big[1+g_B(k_{1},k_{2})\Big(p(k_{1})\left(\hat{k}_{1}\cdot\hat{N}_{A}\right)^2+p(k_{2})\left(\hat{k}_{2}\cdot\hat{N}_{A}\right)^2\nonumber\\\fl&+&p(k_{1})p(k_{2})\left(\hat{k}_{1}\cdot\hat{k}_{2}\right)\left(\hat{k}_{2}\cdot\hat{N}_{A}\right)\left(\hat{k}_{1}\cdot\hat{N}_{A}\right)\Big)\Big]+2\,\,\,perms., 
\eea 
where $\vec{N}_{A}$ is a preferred spatial direction, $B^{iso}$ is an isotropic function and $p$ and $g_B$ are anisotropic coefficients which generally are a function of the moduli of wavevectors (see Eqs.~(\ref{cr}) through (\ref{lcr}) for an explicit definition of these functions). Eq.(\ref{bis1}) is therefore the sum of an isotropic bispectrum plus an anisotropic contribution. For the kind of models we are considering, the isotropic bispectrum ca be taken of the ``local'' form with 
\footnote{For various shapes of primordial non-Gaussianity, see, e.g.~\cite{Cshapes}} 
\bea\label{qui} 
B^{iso}(k_{1},k_{2})=\frac{6}{5} f_{\rm NL} \frac{A^2}{k_{1}^{3}k_{2}^{3}},   
\eea 
where $A$ is the amplitude of the power spectrum of the primordial gravitational potential $P_{\Phi}(k) \propto k^{n-4}$ taken to be scale invariant  (similar definitions hold for $B(k_{1},k_{3})$ and for $B(k_{2},k_{3})$).\\ 

\noindent The bispectrum in Eq.~(\ref{bis1}) will be considered as the primordial source for the bispectrum of temperature fluctuations; we will build an estimator for the parameters that determine the anisotropic correction, treating the bispectrum amplitude in Eq.~(\ref{qui}) as an input of our analysis (see the end of this section for a brief discussion on this).  \\ 

A preliminary computation consists in moving from momentum to harmonic space. In the next few equations, we shortly review these general steps before returning to our specific calculation. \\
The temperature anisotropies are expanded in spherical harmonics 
\bea 
\frac{\Delta T(\hat{n})}{T}=\sum_{lm}a_{lm}Y_{lm}(\hat{n}), 
\eea 
where the $a_{lm}$ coefficients are related to the primordial gravitational potential by 
\bea 
a_{lm}=4\pi(-i)^{l}\int\frac{d^3k}{(2\pi)^3}\Delta_{l}(k)\Phi(\vec{k})Y_{lm}(\hat{k}), 
\eea 
$\Delta_{l}(k)$ being the radiation trasfer function. The bispectrum of temperature fluctuations is  
\bea\fl 
B^{l_{1}\,\,\,\,l_{2}\,\,\,\,l_{3}}_{m_{1}m_{2}m_{3}}\equiv \langle a_{l_{1}m_{1}}a_{l_{2}m_{2}}a_{l_{3}m_{3}} \rangle&=&(4\pi)^3(-i)^{l_{1}+l_{2}+l_{3}}\int\frac{d^3k_{1}}{(2\pi)^3}\frac{d^3k_{2}}{(2\pi)^3}\frac{d^3k_{3}}{(2\pi)^3}\Delta_{l_{1}}(k_{1})\Delta_{l_{2}}(k_{2})\Delta_{l_{3}}(k_{3})\nonumber\\\fl&\times&\langle\Phi(\vec{k}_{1})\Phi(\vec{k}_{2}) \Phi(\vec{k}_{3})\rangle Y_{l_{1}m_{1}}(\hat{k}_{1})Y_{l_{2}m_{2}}(\hat{k}_{2})Y_{l_{3}m_{3}}(\hat{k}_{3}). 
\eea 
Writing the bispectrum of curvature perturbations in the standard way as 
\bea\fl 
\langle\zeta(\vec{k}_{1})\zeta(\vec{k}_{2}) \zeta(\vec{k}_{3})\rangle=(2\pi)^3\delta^{(3)}(\vec{k}_{1}+\vec{k}_{2}+\vec{k}_{3})B(\vec{k}_{1},\vec{k}_{2},\vec{k}_{3})=\left(\frac{5}{3}\right)^3\langle\Phi(\vec{k}_{1})\Phi(\vec{k}_{2}) \Phi(\vec{k}_{3})\rangle 
\eea 
and using 
\bea 
\int d^{3}x e^{i\vec{x}\cdot(\vec{k}_{1}+\vec{k}_{2}+\vec{k}_{3})}=(2\pi)^3\delta^{(3)}(\vec{k}_{1}+\vec{k}_{2}+\vec{k}_{3}), 
\eea 
together with the Rayleigh expansion 
\bea 
e^{i\vec{k}\cdot\vec{x}}=4\pi\sum_{lm}i^{l}j_{l}(kx)Y_{lm}^{*}(\hat{k})Y_{lm}(\hat{x}), 
\eea 
we get 
\bea\label{2}\fl 
B^{l_{1}\,\,\,\,l_{2}\,\,\,\,l_{3}}_{m_{1}m_{2}m_{3}}&=&\left(\frac{3}{5}\right)^3\left(\frac{2}{\pi}\right)^3(-i)^{l_{1}+l_{2}+l_{3}}\int dx x^2\int dk_{1}dk_{2}dk_{3}(k_{1}k_{2}k_{3})^2\Delta_{l_{1}}(k_{1})\Delta_{l_{2}}(k_{2})\Delta_{l_{3}}(k_{3})\nonumber\\\fl&\times&\sum_{l_{1}^{'}m_{1}^{'}}\sum_{l_{2}^{'}m_{2}^{'}}\sum_{l_{3}^{'}m_{3}^{'}}(i)^{l_{1}^{'}+l_{2}^{'}+l_{3}^{'}}j_{l_{1}^{'}}(k_{1}x)j_{l_{2}^{'}}(k_{2}x)j_{l_{3}^{'}}(k_{3}x)\int d\Omega_{\hat{k}_{1}}Y_{l_{1}m_{1}}(\hat{k}_{1})Y_{l_{1}^{'}m_{1}^{'}}^{*}(\hat{k}_{1})\nonumber\\\fl&\times&\int d\Omega_{\hat{k}_{2}}Y_{l_{2}m_{2}}(\hat{k}_{2})Y_{l_{2}^{'}m_{2}^{'}}^{*}(\hat{k}_{2})\int d\Omega_{\hat{k}_{3}}Y_{l_{3}m_{3}}(\hat{k}_{3})Y_{l_{3}^{'}m_{3}^{'}}^{*}(\hat{k}_{3})\times B(\vec{k}_{1},\vec{k}_{2},\vec{k}_{3})\nonumber\\\fl&\times&\int d\Omega_{\hat{x}}Y_{l_{1}^{'}m_{1}^{'}}(\hat{x})Y_{l_{2}^{'}m_{2}^{'}}(\hat{x})Y_{l_{3}^{'}m_{3}^{'}}(\hat{x}). 
\eea 

Let us now replace our parametrization (\ref{bis1}) for the bispectrum. The anisotropic part of the bispectrum from (\ref{bis1}) has a form of this type 
\bea\label{bis00} 
B(\vec{k}_{1},\vec{k}_{2},\vec{k}_{3})\supset B^{iso}\times g_B \times p_{ij}\times\left(\hat{k}_{i}\cdot\hat{k}_{j}\right)\left(\hat{k}_{i}\cdot\hat{N}_{A}\right)\left(\hat{k}_{j}\cdot\hat{N}_{A}\right), 
\eea 
where the indices $i,j$ run over the three wave vectors, $p_{ij}\equiv p(k_{i})$ if $i=j$ and $p_{ij}\equiv p(k_{i})p(k_{j})$ otherwise. The additional assumption we will make at this point in order to simplify our analysis is that the functions $g_B$ and $p$ are actually scale-invariant. This hypothesis is easily met by the models we are dealing with (see e.g. \cite{Bartolo:2009pa}).\\ 
For $i=j$ we can use a simple expansion in spherical harmonics, e.g. 
\bea\label{8} 
p(k_{1})g_B(k_{1},k_{2})\left(\hat{k}_{1}\cdot\hat{N}_{A}\right)^2=\sum_{LM}\lambda_{LM} Y_{LM}(\hat{k}_{1}); 
\eea 
the mixed ($i\neq j$) terms can instead be expressed in terms of bipolar spherical harmonics as, for instance (see, e.g., \cite{Souradeep1,Souradeep2})
\bea\label{88}\fl 
p(k_{1})p(k_{2})g_B(k_{1},k_{2})(\hat{k}_{1}\cdot\hat{k}_{2})\left(\hat{k}_{1}\cdot\hat{N}_{A}\right)\left(\hat{k}_{2}\cdot\hat{N}_{A}\right) 
=\sum_{l_{1},l_{2},L,M}\lambda^{LM}_{l_{1}l_{2}}\{Y_{l_{1}}\bigotimes Y_{l_{2}} \}_{LM}, 
\eea 
where 
\bea 
\label{Ycross}
\{Y_{l_{1}}\bigotimes Y_{l_{2}} \}_{LM}\equiv\sum_{m_{1},m_{2}}C^{LM}_{l_{1}m_{1}l_{2}m_{2}}Y_{l_{1}m_{1}}(\hat{k}_{1})Y_{l_{2}m_{2}}(\hat{k}_{2})
\eea 
and $C^{LM}_{l_{1}m_{1}l_{2}m_{2}}$ are the Clebsch-Gordan coefficients. \\ 

In order to avoid any confusion, throughout the paper we will use upper-case indices $LM$ to characterize the magnitude of statistical anisotropy, 
while the lower-case indices $lm$ will be used for the temperature anisotropies. We will be interested only in coefficients with $L \geq 1$, since $L=0=M$ corresponds to an isotropic monopole 
contribution. In particular, terms like those in Eq.~(\ref{8}) are quadrupolar contributions and, using the orthogonality of the spherical harmonis and the 
spherical addition theorem for the Legendre polynomials, $P_l(\hat{n} \cdot \hat{n}')=(4 \pi/(2l+1)) \sum_{m=-l}^{m=l} Y_{lm}(\hat{n}) Y^*_{lm}(\hat{n}')$, one easily finds 
\begin{equation} 
\label{} 
\lambda_{LM}=\frac{8 \pi}{15}\left(p g_B\right) Y^*_{LM}(\hat{N}_{A}) \delta_{L2}\, , 
\end{equation}   
for $L \geq 1$. However, we will be completely general in using the expansion on the r.h.s. of Eqs.(\ref{8}) and~(\ref{88}) and only at the end of Section~\ref{SNratio} we will specifically apply 
our results for $L=2$, a bispectrum model with a quadrupolar dependence.\\   

We will now write separately the isotropic contribution and the anisotropic ones ((\ref{8}) and (\ref{88})) to $B_{m_{1}m_{2}m_{3}}^{l_{1}\,\,\,\,l_{2}\,\,\,\,l_{3}}$. The isotropic contribution is 
\bea\label{Iso} 
\fl 
B^{l_{1}l_{2}l_{3}(I)}_{m_{1}m_{2}m_{3}}= b_{l_{1}l_{2}l_{3}}G^{l_{1}l_{2}l_{3}}_{m_{1}m_{2}m_{3}}&\equiv& 
\left(\frac{3}{5}\right)^3 \left(\frac{2}{\pi}\right)^3\int dx x^2\int dk_{1}dk_{2}dk_{3}(k_{1}k_{2}k_{3})^2\Delta_{l_{1}}(k_{1})\Delta_{l_{2}}(k_{2})\Delta_{l_{3}}(k_{3})\nonumber\\\fl&\times&j_{l_{1}^{}}(k_{1}x)j_{l_{2}^{}}(k_{2}x)j_{l_{3}^{}}(k_{3}x)B^{iso}(k_{1},k_{2},k_{3}) G^{l_{1}l_{2}l_{3}}_{m_{1}m_{2}m_{3}}\, , 
\eea 
where $b_{l_{1}l_{2}l_{3}}$ is the reduced bipectrum~\cite{Komatsu:2001rj}; the Gaunt integral is given by 
\bea\fl 
G^{l_{1}l_{2}l_{3}}_{m_{1}m_{2}m_{3}}&\equiv&\int d\Omega_{\hat{x}}Y_{l_{1}^{}m_{1}^{}}(\hat{x})Y_{l_{2}^{}m_{2}^{}}(\hat{x})Y_{l_{3}^{}m_{3}^{}}(\hat{x})\nonumber\\\fl&=&\sqrt{\frac{(2l_{1}+1) (2l_{2}+1) (2l_{3}+1)}{4\pi}} 
\left(\begin{array}{ccc}l_1&l_2&l_3\\m_1&m_2&m_3\end{array}\right) 
\left(\begin{array}{ccc}l_1&l_2&l_3\\0&0&0\end{array}\right)
\eea 
and $B^{iso}(k_{1},k_{2},k_{3})\equiv B^{iso}(k_{1},k_{2})+B^{iso}(k_{1},k_{3})+B^{iso}(k_{2},k_{3})$. The sums over $l^{'}_{i}$ and $m^{'}_{i}$ ($i=1,2,3$) were all simplified using the orthogonality condition of the spherical harmonics 
\bea\label{4} 
\int d\Omega_{\hat{k}_{i}}Y_{l_{i}m_{i}}(\hat{k}_{i})Y_{l_{i}^{'}m_{i}^{'}}^{*}(\hat{k}_{i})=\delta_{l_{i}l_{i}^{'}}\delta_{m_{i}m_{i}^{'}}. 
\eea 
The contribution from anisotropic terms like the ones in Eq.~(\ref{8}) instead gives 
\bea\label{5} 
\fl 
B^{l_{1}l_{2}l_{3}(A)}_{m_{1}m_{2}m_{3}}&=&\left(\frac{3\times 6}{5\pi}\right)^3 \sum_{l_{1}^{'}m_{1}^{'}}\sum_{LM} 
\lambda_{LM} G^{l_{1}^{'}l_{2}l_{3}}_{m_{1}^{'}m_{2}m_{3}}G^{l_{1}l_{1}^{'}L_{}}_{m_{1}-m_{1}^{'}M_{}}(-1)^{l_{1}}(i)^{l_{1}+l^{'}_{1}}(-1)^{m^{'}_{1}} 
b_{l_{1}l_{2}l_{3}}^{l^{'}_{1}}\nonumber\\\fl 
&+&\left(\frac{3\times 6}{5\pi}\right)^3 \sum_{l_{2}^{'}m_{2}^{'}}\sum_{LM} 
\lambda_{LM} G^{l_{1}^{}l_{2}^{'}l_{3}}_{m_{1}^{}m_{2}^{'}m_{3}}G^{l_{2}l_{2}^{'}L_{}}_{m_{2}-m_{2}^{'}M_{}}(-1)^{l_{2}}(i)^{l_{2}+l^{'}_{2}}(-1)^{m^{'}_{2}} 
b_{l_{1}l_{2}l_{3}}^{\,\,\,\,\,l^{'}_{2}}\nonumber\\\fl 
&+&\left(\frac{3\times 6}{5\pi}\right)^3 \sum_{l_{3}^{'}m_{3}^{'}}\sum_{LM} 
\lambda_{LM} G^{l_{1}^{}l_{2}l_{3}^{'}}_{m_{1}^{}m_{2}m_{3}^{'}}G^{l_{3}l_{3}^{'}L_{}}_{m_{3}-m_{3}^{'}M_{}}(-1)^{l_{3}}(i)^{l_{3}+l^{'}_{3}}(-1)^{m^{'}_{1}} 
b_{l_{1}l_{2}l_{3}}^{\,\,\,\,\,\,\,\,\,l^{'}_{3}},\nonumber\\\fl 
\eea 
where 
\bea 
\label{b1} 
\fl 
b_{l_{1}l_{2}l_{3}}^{l_{1}^{'}}&\equiv&\int dx x^2\int dk_{1}dk_{2}dk_{3}(k_{1}k_{2}k_{3})^2\Delta_{l_{1}}(k_{1})\Delta_{l_{2}}(k_{2})\Delta_{l_{3}}(k_{3})\nonumber\\\fl&\times&j_{l_{1}^{'}}(k_{1}x)j_{l_{2}^{}}(k_{2}x)j_{l_{3}^{}}(k_{3}x)\left(B^{iso}(k_{1},k_{2})+B^{iso}(k_{1},k_{3})\right), 
\eea 
\bea 
\label{b2} 
\fl 
b_{l_{1}l_{2}l_{3}}^{\,\,\,\,\,l^{'}_{2}}&\equiv&\int dx x^2\int dk_{1}dk_{2}dk_{3}(k_{1}k_{2}k_{3})^2\Delta_{l_{1}}(k_{1})\Delta_{l_{2}}(k_{2})\Delta_{l_{3}}(k_{3})\nonumber\\\fl&\times&j_{l_{1}^{}}(k_{1}x)j_{l_{2}^{'}}(k_{2}x)j_{l_{3}^{}}(k_{3}x)\left(B^{iso}(k_{1},k_{2})+B^{iso}(k_{2},k_{3})\right), 
\eea 
\bea 
\label{b3} 
\fl 
b_{l_{1}l_{2}l_{3}}^{\,\,\,\,\,\,\,\,\,l^{'}_{3}}&\equiv&\int dx x^2\int dk_{1}dk_{2}dk_{3}(k_{1}k_{2}k_{3})^2\Delta_{l_{1}}(k_{1})\Delta_{l_{2}}(k_{2})\Delta_{l_{3}}(k_{3})\nonumber\\\fl&\times&j_{l_{1}^{}}(k_{1}x)j_{l_{2}^{}}(k_{2}x)j_{l_{3}^{'}}(k_{3}x)\left(B^{iso}(k_{1},k_{3})+B^{iso}(k_{2},k_{3})\right). 
\eea 
Eq.~(\ref{4}) was employed for $i=2,3$, whereas for $i=1$ we have an integral over three spherical harmonics which produces the extra (compared to the isotropic case) Gaunt integral appearing in Eq.~(\ref{5}).\\ 
Finally, let us consider the anisotropic terms like those in (\ref{88}) 
\bea\label{6} 
\fl 
B^{l_{1}\,\,\,\,l_{2}\,\,\,\,l_{3}}_{m_{1}m_{2}m_{3}}&\supset&\left(\frac{3}{5}\right)^3 \left(\frac{2}{\pi}\right)^3\sum_{LMl^{'}l^{''}m^{'}m^{''}}\sum_{l^{'}_{1}m^{'}_{1}l^{'}_{2}m^{'}_{2}}\lambda^{LM}_{l^{'}l^{''}} C_{l^{'}m^{'}l^{''}m^{''}}^{LM} G^{l^{'}_{1}l^{'}_{2}l_{3}}_{m^{'}_{1}m^{'}_{2}m_{3}}G^{l_{1}l_{1}^{'}l^{'}}_{m_{1}-m_{1}^{'}m^{'}}G^{l^{}_{2}l^{'}_{2}l^{''}}_{m_{2}-m_{2}^{'}m^{''}}\nonumber\\\fl&\times& 
(-1)^{m^{'}_{1}+m^{'}_{2}}(-1)^{l_{1}+l_{2}}(i)^{l^{}_{1}+l_{1}^{'}}(i)^{l^{}_{2}+l_{2}^{'}}b_{l_{1}l_{2}l_{3}}^{l_{1}^{'}l_{2}^{'}}+\,\, perms., 
\eea 

where 
\bea\fl 
b_{l_{1}l_{2}l_{3}}^{l_{1}^{'}l_{2}^{'}}&\equiv&\int dx x^2\int dk_{1}dk_{2}dk_{3}(k_{1}k_{2}k_{3})^2\Delta_{l_{1}}(k_{1})\Delta_{l_{2}}(k_{2})\Delta_{l_{3}}(k_{3})\nonumber\\\fl&\times&j_{l_{1}^{'}}(k_{1}x)j_{l_{2}^{'}}(k_{2}x)j_{l_{3}^{}}(k_{3}x)\left(B^{iso}(k_{1},k_{2})+B^{iso}(k_{1},k_{3})\right). 
\eea 
 
Notice that in the vector field models we are considering, the isotropic bispectrum is \textsl{separable} (remember Eq.~(\ref{qui})), i.e. we can introduce three functions $X$, $Y$ and $Z$ 
\bea\fl 
(k_{1}k_{2}k_{3})^2 B^{iso}(k_{1},k_{2},k_{3})=X(k_{1})Y(k_{2})Z(k_{3})+perms. 
\eea 
that give 
\bea\label{7} 
b_{l_{1}l_{2}l_{3}}=\int dx x^2 X_{l_{1}}(k_{1})Y_{l_{2}}(k_{2})Z_{l_{3}}(k_{3})+\,\, perms.,\\ 
b_{l_{1}l_{2}l_{3}}^{l_{1}^{'}}=  \int dx x^2 X_{l_{1}}^{l^{'}_{1}}(k_{1})Y_{l_{2}}(k_{2})Z_{l_{3}}(k_{3})+\,\, perms.,\\ 
b_{l_{1}l_{2}l_{3}}^{l_{1}^{'}l_{2}^{'}}=\int dx x^2 X_{l_{1}}^{l^{'}_{1}}(k_{1})Y_{l_{2}}^{l^{'}_{2}}(k_{2})Z_{l_{3}}(k_{3})+\,\, perms., 
\eea 
where 
\bea 
X_{l_{1}}(x)\equiv \int dk_{1}\Delta_{l_{1}}(k_{1})X(k_{1})j_{l_{1}}(k_{1}x)  ,\\ 
X_{l_{1}}^{l^{'}_{1}}(k_{1})\equiv \int dk_{1}\Delta_{l_{1}}(k_{1})X(k_{1})j_{l_{1}^{'}}(k_{1}x)    , 
\eea 
and so on for the other integrals in (\ref{7}). \\ 

We will now further simplify our analysis by considering an anisotropic contribution as in Eq.~(\ref{8}) only. We have checked that the anisotropic terms as in Eq.~(\ref{88}), for $L>0$, contain either contributions that can be recast as the angular decomposition of  Eq.~(\ref{8}) or contributions that display an angular dependence that is actually different from the one of 
 Eq.~(\ref{8}) and which would correspond to a different anisotropic signature. Therefore a distinction between these two contributions can be drawn and from now on we focus only on an analysis of  the  $\lambda_{\rm LM}$s coefficients of the anisotropic bispectrum. This means that our results will be targeted only to the anisotropic features of the kind of Eq.~(\ref{8}). A similar dedicated analysis can be performed to other types of bispectrum anisotropic features.\footnote{
We have studied in some details the angular dependence of the l.h.s. of Eq.~(\ref{88}). We have seen that a quadrupole bispectrum anisotropy is generated, with $L=2$. As an example of what we mentioned above, it is easy to realize that, looking at Eqs.~(\ref{88}) and~(\ref{Ycross}), for $(l_1=0,l_2=2)$, or for $(l_1=2,l_2=0)$ one  just recovers an angular dependence as in Eq.~(\ref{8}), while  the other remaining combination ($l_1=l_2=2$) displays a different anisotropic signature. 

Notice also that we have found that Eq.~(\ref{88}) (multiplied by $B^{iso}(k_1,k_2)$ appearing in Eq.~(\ref{bis1})) for $L=0$ contains an additional isotropic bispectrum. We have carefully checked that such an isotropic bispectrum has a significant correlation (more than $72\%$)  with $B^{iso}$ (using the correlation defined in~\cite{Cshapes}). In this way, when we factor out the isotropic bispectrum $B^{iso}$ in Eq.~(\ref{bis1}), the effect of this additional isotropic bispectrum from the second line of Eq.~(\ref{bis1}) is to produce with reasonable approximation a constant term that can be reabsorbed as described later in Eqs.~(\ref{r1}) 
and~(\ref{r2}).} Therefore, let us write the  bispectrum, given by the sum of an isotropic and an anisotropic part, as 

\bea\label{qq} 
B_{m_{1}m_{2}m_{3}}^{l_{1}\,\,\,\,l_{2}\,\,\,\,l_{3}}&\equiv&\langle a_{l_{1}m_{1}}a_{l_{2}m_{2}}a_{l_{3}m_{3}} \rangle=B_{m_{1}m_{2}m_{3}}^{l_{1}l_{2}l_{3}(I)}+ B_{m_{1}m_{2}m_{3}}^{l_{1}l_{2}l_{3}(A)}\nonumber\\&=&f_{NL}\left(B_{m_{1}m_{2}m_{3}}^{l_{1}l_{2}l_{3}(I)}|_{f_{NL}=1}+\sum_{LM}\lambda_{LM} B_{m_{1}m_{2}m_{3}}^{l_{1}l_{2}l_{3}(A)LM}|_{f_{NL}=1}\right), 
\eea 
where $B_{m_{1}m_{2}m_{3}}^{l_{1}l_{2}l_{3}(I)}$ and $B_{m_{1}m_{2}m_{3}}^{l_{1}l_{2}l_{3}(A)LM}$ can be read off Eqs.~(\ref{Iso}) and (\ref{5}) respectively. We have factorized the (isotropic) non-Gaussianity amplitude $f_{NL}$, defined in the traditional way as proportional to the ratio between the (isotropic) bispectrum and the (isotropic) power spectrum squared in the equilateral configuration.\\ 

Estimators for the $\lambda_{LM}$'s will be computed in the next Section. Before proceeding though, it is necessary to make some further considerations. \\

First of all, for the rest of the paper, we will select vector field models where the bispectrum of curvature fluctuations is real, which is equivalent to assuming parity conservation in our theories; indeed, it is easy to realize that $\langle\zeta_{\vec{k}_{1}}^{*}\zeta_{\vec{k}_{2}}^{*}\zeta_{\vec{k}_{3}}^{*}\rangle=\langle \zeta_{-\vec{k}_{1}}^{}\zeta_{-\vec{k}_{2}}^{}\zeta_{-\vec{k}_{3}}^{}\rangle$, therefore imposing the reality of the bispectrum corresponds exactly to require that
\bea
\langle \zeta_{\vec{k}_{1}}^{}\zeta_{\vec{k}_{2}}^{}\zeta_{\vec{k}_{3}}^{}\rangle=\langle \zeta_{-\vec{k}_{1}}^{}\zeta_{-\vec{k}_{2}}^{}\zeta_{-\vec{k}_{3}}^{}\rangle,
\eea
i.e.
\bea 
B(-\vec{k}_{1},-\vec{k}_{2},-\vec{k}_{3})=B(\vec{k}_{1},\vec{k}_{2},\vec{k}_{3})
\eea
which, (using Eq.~(\ref{5})), is only satisfied if 
\bea\label{evenL}
L=even.
\eea
Also, consider the well-known property $a^{*}_{lm}=(-1)^{m}a_{l-m}$, then 
\bea
\left(B_{m_{1}m_{2}m_{3}}^{l_{1}\,\,\,\,l_{2}\,\,\,\,l_{3}}\right)^{*}=(-1)^{m_{1}+m_{2}+m_{3}}B_{-m_{1}-m_{2}-m_{3}}^{\,\,l_{1}\,\,\,\,\,\,l_{2}\,\,\,\,\,\,l_{3}}. 
\eea
The latter equation, together with (\ref{evenL}), are only satisfied if 
\bea\label{p11} 
\lambda_{LM}^{*}=(-1)^{M}\lambda_{L-M}.
\eea 
Notice also that if (\ref{evenL}) and (\ref{p11}) are true, then $B_{m_{1}m_{2}m_{3}}^{l_{1}l_{2}l_{3}(A)LM}$ is real. \\

With Eq.~(\ref{evenL}) in mind, we can now split the sum on the r.h.s. of Eq.~(\ref{8}) as follows
\bea
\sum_{LM}\lambda_{LM}Y_{LM}(\hat{k})=\lambda_{00}Y_{00}+\sum_{L\geq 2, M}\lambda_{LM}Y_{LM}(\hat{k}).
\eea
The first term of the r.h.s. is a constant (remember that we are assuming no momentum dependence for the anisotropy parameters), so it can be conveniently incorporated into $B^{iso}$ through a rescaling
\bea
\label{r1}
B^{iso}\rightarrow B^{iso}(1+2\lambda_{00}Y_{00});
\eea
similarly, the anisotropy parameters would be rescaled
\bea
\label{r2}
\lambda_{LM}\rightarrow\frac{\lambda_{LM}}{1+2\lambda_{00}Y_{00}}.
\eea 
With these rescalings, the $L=0$ contribution will be included in $B_{m_{1}m_{2}m_{3}}^{l_{1}l_{2}l_{3}(I)}$ and so the sum in Eq.~(\ref{qq}) will be from now on intended over $L\geq 2$. \\

Finally, we would like to quickly comment on the a priori determination of $f_{NL}$ that, as previously mentioned, we will treat as an input parameter while estimating the $\lambda_{LM}$'s. $f_{NL}$ can be estimated from the angle averaged bispectrum defined as
\bea
B_{l_{1}l_{2}l_{3}}\equiv \sum_{m_{1},m_{2},m_{3}}\left(\begin{array}{ccc}l_{1}^{}&l_{2}&l_{3}\\m_{1}&m_{2}&m_{3}\end{array}\right)B_{m_{1}m_{2}m_{3}}^{l_{1}l_{2}l_{3}},
\eea
to which the anisotropic part provides, as expected, a vanishing contribution 
\bea\label{uff}\fl
\sum_{m_{1},m_{2},m_{3}}\left(\begin{array}{ccc}l_{1}^{}&l_{2}&l_{3}\\m_{1}&m_{2}&m_{3}\end{array}\right)B_{m_{1}m_{2}m_{3}}^{l_{1}l_{2}l_{3}(A)LM}&\simeq& \Pi_{l_{1}l_{2}l_{3}}\Big(c_{l_{1}}b_{l_{1}l_{2}l_{3}}^{l_{1}}+c_{l_{2}}b_{l_{1}l_{2}l_{3}}^{\,\,\,\,\,l_{2}}+c_{l_{3}}b_{l_{1}l_{2}l_{3}}^{\,\,\,\,\,\,\,\,\,\,l_{3}}\Big)\delta_{L0}\delta_{M0}=0\nonumber\\.
\eea
(using Eq.~(\ref{5}) and $L\geq 2$). In (\ref{uff}) We define $c_{l_{i}}\equiv \sqrt{2l_{i}+1}$ ($i=1,2,3$) and $\Pi_{l_{1}l_{2}l_{3}}\equiv \sqrt{(2l_{1}+1)(2l_{2}+1)(2l_{3}+1)}$. This last result accounts for the possibility of determining the isotropic non-Gaussianity amplitude independently from the anisotropic one. In fact, the main reason why we 
perform the rescaling~(\ref{r1}) and~(\ref{r2}) is exactly that an estimator measuring the isotropic $f_{\rm NL}$ from the angle-averaged bispectrum would eventually also pick up the monopole contribution $L=0$, and so we must take it into account when defining the isotropic part of the bispectrum.


\section{Estimator for statistical anisotropy from the bispectrum} 
In this section we will derive an estimator for the amplitudes of statistical anisotropy $\lambda_{LM}$. We start with 
the expression of the probability distribution function (PDF) for a general non-Gaussian case employing the Edgeworth expansion \cite{Bernardeau:1994aq,Blinnikov:1997jq,Taylor:2000hq,Babich:2005en} 
\bea\label{1o}\fl 
P(a)=\left(1-\frac{1}{6}\sum_{l_{i},m_{i}}\langle a_{l_{1}m_{1}}a_{l_{2}m_{2}}a_{l_{3}m_{3}} \rangle\frac{\p}{\p a_{l_{1}m_{1}}}\frac{\p}{\p a_{l_{2}m_{2}}}\frac{\p}{\p a_{l_{3}m_{3}}}+...\right)\frac{e^{-\frac{1}{2}\sum_{l_{q},m_{q}}a^{*}_{l_{4}m_{4}}C^{-1}_{l_{4}m_{4},l_{5}m_{5}}a_{l_{5}m_{5}}}}{\left(2\pi\right)^{N_{p}/2}\left(detC\right)^{1/2}},\nonumber\\ 
\eea 
where $i=1,2,3$, $q=4,5$, $N_{p}$ is the number of pixels of a given measurement and $C$ is the covariance matrix 
\bea 
C_{l_{}m_{},l^{'}m^{'}}\equiv\langle a_{l_{}m_{}}a^{*}_{l^{'}m^{'}}\rangle. 
\eea 
We now truncate the Edgeworth expansion at third order: one can verify that these are all the terms we need for the computation of the estimators and of the signal-to-noise ratios of the $\lambda_{LM}$s coefficients, if we make the choice to work at leading order in $f_{NL}$ and also to neglect all connected correlators of order larger than three (see Appendix~C for more details).\\
Using $a_{lm}^{*}=(-1)^{m}a_{l-m}$ and the relations 
\bea 
C_{l-m,l^{'}-m^{'}}=(-1)^{m+m^{'}}C_{l^{'}m^{'},lm},\\ 
\left(C_{lm,l^{'}m^{'}}\right)^{*}=C_{l^{'}m^{'},lm}, 
\eea 
which also apply to the inverse matrix 
\bea 
C^{-1}_{l-m,l^{'}-m^{'}}=(-1)^{m+m^{'}}C^{-1}_{l^{'}m^{'},lm},\\ 
\left(C^{-1}_{lm,l^{'}m^{'}}\right)^{*}=C^{-1}_{l^{'}m^{'},lm}, 
\eea 
Eq.~(\ref{1o}) becomes 
\bea\label{3o}\fl 
P(a)&=&\Big[1+\frac{1}{6}\sum_{l_{i},m_{i}=1,2,3}\langle a_{l_{1}m_{1}}a_{l_{2}m_{2}}a_{l_{3}m_{3}} \rangle\sum_{l_{q},m_{q}=4,5,6}\Big((-1)^{m_{1}+m_{2}+m_{3}}C^{-1}_{l_{1}-m_{1},l_{4}m_{4}}a_{l_{4}m_{4}}\\\fl&\times&C^{-1}_{l_{2}-m_{2},l_{5}m_{5}}a_{l_{5}m_{5}}C^{-1}_{l_{3}-m_{3},l_{6}m_{6}}a_{l_{6}m_{6}}-(-1)^{m_{1}+m_{2}}C^{-1}_{l_{2}-m_{2},l_{3}m_{3}}C^{-1}_{l_{1}-m_{1},l_{4}m_{4}}a_{l_{4}m_{4}}\nonumber\\\fl&-&(-1)^{m_{1}+m_{2}}C^{-1}_{l_{1}-m_{1},l_{3}m_{3}}C^{-1}_{l_{2}-m_{2},l_{4}m_{4}}a_{l_{4}m_{4}}-(-1)^{m_{1}+m_{3}}C^{-1}_{l_{1}-m_{1},l_{2}m_{2}}C^{-1}_{l_{3}-m_{3},l_{4}m_{4}}a_{l_{4}m_{4}}\Big)\Big]\nonumber\\\fl&\times&\frac{e^{-\frac{1}{2}\sum_{l_{q},m_{q}}a^{*}_{l_{4}m_{4}}C^{-1}_{l_{4}m_{4},l_{5}m_{5}}a_{l_{5}m_{5}}}}{\left(2\pi\right)^{N_{p}/2}\left(detC\right)^{1/2}}\nonumber. 
\eea 
Let us now consider a diagonal variance. This choice is motivated by the fact that the non-diagonal part of the variance would generally introduce a correction proportional to the statistical anisotropy amplitude of the power spectrum (see e.g. \cite{Hanson:2009gu}) which, as previously mentioned, is known to be quite small both from experiments and in the theoretical models we are referring to. \footnote{
Indeed a non-diagonal variance can be generated by anisotropies generated by various experimental effects (such as masking or inhomogeneous noise). These effects can be treated 
with the linear term widely used in the bispectrum analyses, see, e.g., Ref.~\cite{CNTZ}.} 

Eq.~(\ref{3o}) will then reduce to 
\bea\label{4o}\fl 
P(a)&=&\Big[1+\frac{1}{6}\sum_{l_{i},m_{i}=1,2,3}\langle a_{l_{1}m_{1}}a_{l_{2}m_{2}}a_{l_{3}m_{3}} \rangle\Big(\frac{a^{*}_{l_{1}m_{1}}a^{*}_{l_{2}m_{2}}a^{*}_{l_{3}m_{3}}}{C_{l_{1}}C_{l_{2}}C_{l_{3}}}-\frac{(-1)^{m_{2}}}{C_{l_{1}}C_{l_{2}}}\delta_{l_{2}l_{3}}\delta_{m_{2}-m_{3}}a^{*}_{l_{1}m_{1}}\nonumber\\\fl&-&\frac{(-1)^{m_{1}}}{C_{l_{1}}C_{l_{2}}}\delta_{l_{1}l_{3}}\delta_{m_{1}-m_{3}}a^{*}_{l_{2}m_{2}}-\frac{(-1)^{m_{1}}}{C_{l_{1}}C_{l_{3}}}\delta_{l_{1}l_{2}}\delta_{m_{1}-m_{2}}a^{*}_{l_{3}m_{3}}\Big)\Big]\times\frac{e^{-\frac{1}{2}\sum_{l,m}a^{*}_{l_{}m_{}}C^{-1}_{l}a_{l_{}m_{}}}}{\left(2\pi\right)^{N_{p}/2}\left(detC\right)^{1/2}}.\nonumber\\\fl 
\eea 
It is easy to verify that the linear part of Eq.~(\ref{4o}) is proportional to the monopole in the isotropic case. Let us for instance consider the first linear term, i.e. proportional to $a^{*}_{l_{1}m_{1}}$: from $\delta_{m_{2}-m_{3}}$ and from $m_{1}+m_{2}+m_{3}=0$ (which is always satisfied if the bispectrum is isotropic) we have $m_{1}=0$; using the identity in Eq.~(\ref{10o}), we get $l_{1}=0$, so the only nonzero contribution is due to $a^{*}_{00}$. A similar reasoning can be applied to the second and the third linear contributions in Eq.~(\ref{4o}). 
This does not apply however to the anisotropic case, so it is necessary to retain the linear part of the expression.\\ 

We are now ready to compute the estimator for the anisotropy parameters $\lambda_{LM}$ using the expression for the total bispectrum, Eq.~(\ref{qq}). 
For small non-Gaussianity and for small statistical anisotropy, the PDF can be expanded as 
\bea\label{pdfexp} 
\ln(P)&=&\ln(1+x)-\frac{1}{2}\sum_{l,m}a^{*}_{l_{}m_{}}C^{-1}_{l}a_{l_{}m_{}}+const.\nonumber\\&\simeq& x-\frac{x^2}{2}-\frac{1}{2}\sum_{l,m}a^{*}_{l_{}m_{}}C^{-1}_{l}a_{l_{}m_{}}+const., 
\eea 
where 
\bea\label{xdef}\fl 
x&\equiv& \frac{f_{NL}}{6}\sum_{l_{i}m_{i}}\left[B_{m_{1}m_{2}m_{3}}^{l_{1}l_{2}l_{3}(I)}|_{f_{NL}=1}+\sum_{LM}\lambda_{LM} B_{m_{1}m_{2}m_{3}}^{l_{1}l_{2}l_{3}(LM)}|_{f_{NL}=1}\right]\Big(\frac{a^{*}_{l_{1}m_{1}}a^{*}_{l_{2}m_{2}}a^{*}_{l_{3}m_{3}}}{C_{l_{1}}C_{l_{2}}C_{l_{3}}}\nonumber\\\fl&-&\frac{(-1)^{m_{2}}}{C_{l_{1}}C_{l_{2}}}\delta_{l_{2}l_{3}}\delta_{m_{2}-m_{3}}a^{*}_{l_{1}m_{1}}-\frac{(-1)^{m_{1}}}{C_{l_{1}}C_{l_{2}}}\delta_{l_{1}l_{3}}\delta_{m_{1}-m_{3}}a^{*}_{l_{2}m_{2}}-\frac{(-1)^{m_{1}}}{C_{l_{1}}C_{l_{3}}}\delta_{l_{1}l_{2}}\delta_{m_{1}-m_{2}}a^{*}_{l_{3}m_{3}}\Big). 
\eea 
Let us introduce the more compact notation 
\bea\fl 
B^{(I)}&\equiv&\frac{1}{6}\sum_{l_{i}m_{i}}B_{m_{1}m_{2}m_{3}}^{l_{1}l_{2}l_{3}(I)}|_{f_{NL}=1}\Big(\frac{a^{*}_{l_{1}m_{1}}a^{*}_{l_{2}m_{2}}a^{*}_{l_{3}m_{3}}}{C_{l_{1}}C_{l_{2}}C_{l_{3}}}-\frac{(-1)^{m_{2}}}{C_{l_{1}}C_{l_{2}}}\delta_{l_{2}l_{3}}\delta_{m_{2}-m_{3}}a^{*}_{l_{1}m_{1}}\nonumber\\\fl&-&\frac{(-1)^{m_{1}}}{C_{l_{1}}C_{l_{2}}}\delta_{l_{1}l_{3}}\delta_{m_{1}-m_{3}}a^{*}_{l_{2}m_{2}}-\frac{(-1)^{m_{1}}}{C_{l_{1}}C_{l_{3}}}\delta_{l_{1}l_{2}}\delta_{m_{1}-m_{2}}a^{*}_{l_{3}m_{3}}\Big),\\\fl 
B^{(A)}&\equiv& \lambda_{LM} B^{(A) LM} \equiv \frac{1}{6}\lambda_{LM} \sum_{l_{i}m_{i}}B_{m_{1}m_{2}m_{3}}^{l_{1}l_{2}l_{3}(A) LM}|_{f_{NL}=1}\Big(\frac{a^{*}_{l_{1}m_{1}}a^{*}_{l_{2}m_{2}}a^{*}_{l_{3}m_{3}}}{C_{l_{1}}C_{l_{2}}C_{l_{3}}}\\\fl&-&\frac{(-1)^{m_{2}}}{C_{l_{1}}C_{l_{2}}}\delta_{l_{2}l_{3}}\delta_{m_{2}-m_{3}}a^{*}_{l_{1}m_{1}}-\frac{(-1)^{m_{1}}}{C_{l_{1}}C_{l_{2}}}\delta_{l_{1}l_{3}}\delta_{m_{1}-m_{3}}a^{*}_{l_{2}m_{2}}-\frac{(-1)^{m_{1}}}{C_{l_{1}}C_{l_{3}}}\delta_{l_{1}l_{2}}\delta_{m_{1}-m_{2}}a^{*}_{l_{3}m_{3}}\Big),\nonumber 
\eea 
where  sum over $(L,M)$ is understood. Notice that the operators $B^{(I)}$ and $B^{(A)}$ are real and commute with each other. So 
\bea\fl 
\ln(P)&\simeq& f_{NL}\left[B^{(I)}+\lambda_{LM} B^{(A)LM}\right]-\frac{f_{NL}^{2}}{2}\left[\left(B^{(I)}\right)^2+2 B^{(I)}\lambda_{LM} B^{(A)LM}+\left(\lambda_{LM} B^{(A)LM}\right)^2\right]\nonumber\\\fl&-&\frac{1}{2}\sum_{l,m}a^{*}_{l_{}m_{}}C^{-1}_{l}a_{l_{}m_{}}+const. 
\eea 
\noindent Let us call $\hat{\lambda}_{LM}$ our estimator for the anisotropy parameters. We require $\hat{\lambda}_{LM}$ to be unbiased and optimal estimators
\bea\label{unb} 
\langle \hat{\lambda}_{LM}\rangle&=&\lambda_{LM},\\\label{opt} 
\sigma^2_{\lambda_{LM}}&=&\left(F^{-1}\right)_{\lambda_{LM}\lambda_{LM}}, 
\eea 
where $F$ is the Fisher matrix whose $(LM,L'M')$ entry is defined as \cite{Hanson:2009gu, ingegneri} (see Appendix B for more details) 
\bea\label{fmLM} 
F_{\lambda_{LM}\lambda_{L'M'}}\equiv-\Bigg\langle\frac{\p^2 \ln(P)}{\p \lambda_{LM}\p\lambda^{*}_{L'M'}} \Bigg\rangle=\Bigg\langle\left(\frac{\p \ln(P)}{\p \lambda_{LM}}\right)\left(\frac{\p \ln(P)}{\p \lambda_{L'M'}^{*}}\right) \Bigg\rangle. 
\eea   
It is possible to prove that, if the Fisher matrix is diagonal (which is our case as we show in the next section) 
a necessary and sufficient condition for an estimator $\hat{\lambda}_{LM}$ to be optimal is the following (see for example~\cite{Babich:2005en}) 
\bea\label{ii} 
\frac{\p \ln(P)}{\p\lambda_{LM}}=F_{\lambda_{LM}\lambda_{LM}}\left(\hat{\lambda}_{LM}-\lambda_{LM}\right). 
\eea 
We will assume that all other cosmological parameters, including $f_{NL}$, are known beforehand and, from there, determine the estimator for $\lambda_{LM}$. 
The Fisher matrix is 
\bea\label{qq1} 
F_{\lambda_{LM}\lambda_{LM}}=f_{NL}^2\Bigg\langle B^{(A)LM}  {B^{(A)LM}}^*\Bigg\rangle \, . 
\eea 
One can check that Eq.~(\ref{ii}) is satisfied by 
\bea\label{qq2} 
\hat{\lambda}_{LM}=\frac{1}{F_{\lambda_{LM}\lambda_{LM}}}\left(f_{NL}B^{(A)LM}-f_{NL}^{2}B^{(I)}B^{(A) LM}\right)\, , 
\eea 
if in the expression for $\left(\partial \ln P / \partial \lambda_{LM}\right)$ 
\begin{equation}\label{derp} 
\frac{\partial \ln P}{\partial \lambda_{LM}}=f_{NL} B^{(A)LM} -f_{NL}^2 B^{(I)} B^{(A)LM}-f_{NL}^2 \lambda^*_{L'M'} {B^{(A)L'M' }}^* B^{(A)LM}\, , 
\end{equation} 
one replaces the last product with its expectation value, ${B^{(A)L'M'}}^* B^{(A)LM} \rightarrow  \langle {B^{(A)L'M'}}^* B^{(A)LM} \rangle$. This approximation is justified in the case of a low level 
of statistical anisotropy (and small non-Gaussianity). Therefore, in this limit, the estimator is given by Eq.~(\ref{qq2}). The first term on the left-hand side of (\ref{qq2}) has a familiar structure if one considers for instance the expression for the estimator of the non-Gaussianity amplitude $f_{NL}$; however, in this case, we expect the subtraction of a term proportional to the isotropic part ($B^{(I)}$) of the bispectrum, since all of the information about statistical anisotropy is encoded in $B^{(A)LM}$. Notice also that the second term is \textsl{not} subleading compared to first term as the second power of $f_{NL}$ might apparently suggest. The Fisher matrix $F_{\lambda_{LM}\lambda_{LM}}$ will be computed in the next section.\\


\section{Signal-to-noise ratio} 
\label{SNratio} 

The Fisher matrix is a very powerful piece of information since it sets a lower bound on experimental errors via the Cramer-Rao inequality (see, e.g., 
\cite{Babich:2005en,Hamilton2005,Verde:2009tu,Heavens:2009nx,Liguori:2010hx,Kamion} and references therein). In fact our estimator~(\ref{}) is optimal in the sense 
that the Cramer-Rao inequality is saturated, that is the variance of the estimator equals the one given by the Fisher matrix 
\begin{equation} 
\sigma^2_{\lambda_{LM}}=\left( F^{-1} \right)_{\lambda_{LM}\lambda_{LM}}\, . 
\end{equation} 
The signal-to-noise ratios for our parameters $\lambda_{LM}$ are therefore 
\bea 
\left(\frac{S}{N}\right)_{\lambda_{LM}}=\frac{\lambda_{LM}}{\sqrt{\left(F^{-1}\right)_{\lambda_{LM}\lambda_{LM}}}}. 
\eea 
Let us now compute the Fisher matrix as given in Eq.(\ref{fmLM}). We will compute it in the limit where the variance is diagonal and to leading order in $f_{NL}$. Is it convenient to report the complete expression for the anisotropic contribution to the bispetrum from Sec.~3 
\bea\label{ba}\fl 
B^{(A)}&\equiv&\frac{1}{6}\lambda_{LM} \sum_{l_{i}m_{i}} B_{m_{1}m_{2}m_{3}}^{l_{1}l_{2}l_{3}(A) LM}|_{f_{NL}=1}\Big(\frac{a^{*}_{l_{1}m_{1}}a^{*}_{l_{2}m_{2}}a^{*}_{l_{3}m_{3}}}{C_{l_{1}}C_{l_{2}}C_{l_{3}}}-\frac{(-1)^{m_{2}}}{C_{l_{1}}C_{l_{2}}}\delta_{l_{2}l_{3}}\delta_{m_{2}-m_{3}}a^{*}_{l_{1}m_{1}}\nonumber\\\fl&-&\frac{(-1)^{m_{1}}}{C_{l_{1}}C_{l_{2}}}\delta_{l_{1}l_{3}}\delta_{m_{1}-m_{3}}a^{*}_{l_{2}m_{2}}-\frac{(-1)^{m_{1}}}{C_{l_{1}}C_{l_{3}}}\delta_{l_{1}l_{2}}\delta_{m_{1}-m_{2}}a^{*}_{l_{3}m_{3}}\Big)\, , 
\eea 
where a sum over $(L, M)$ is understood. As we can see from the previous equation, $10$ contributions arise for $F_{\lambda_{LM}\lambda_{LM}}$: $1$ from the six point function $\langle a^{*}_{l_{1}m_{1}}a^{*}_{l_{2}m_{2}}a^{*}_{l_{3}m_{3}}a_{l_{4}m_{4}}a_{l_{5}m_{5}}a_{l_{6}m_{6}}\rangle$, $6$ from the four point functions $\langle a^{*}_{l_{1}m_{1}}a^{*}_{l_{2}m_{2}}a^{*}_{l_{3}m_{3}}a_{l_{4}m_{4}}\rangle$ and $9$ from the two point function $\langle a^{*}_{l_{1}m_{1}} a_{l_{4}m_{4}}\rangle$. We provide their expressions below 
\bea\fl 
\langle a^{*}_{l_{1}m_{1}}a^{*}_{l_{2}m_{2}}a^{*}_{l_{3}m_{3}}a_{l_{4}m_{4}}a_{l_{5}m_{5}}a_{l_{6}m_{6}}\rangle&=&\delta_{l_{1}l_{4}}\delta_{l_{2}l_{5}}\delta_{l_{3}l_{6}}\delta_{m_{1}m_{4}}\delta_{m_{2}m_{5}}\delta_{m_{3}m_{6}}C_{l_{1}}C_{l_{2}}C_{l_{3}}\nonumber\\&+&\delta_{l_{1}l_{2}}\delta_{l_{3}l_{4}}\delta_{l_{5}l_{6}}\delta_{m_{1}-m_{2}}\delta_{m_{3}m_{4}}\delta_{m_{5}-m_{6}}(-1)^{m_{1}+m_{5}}C_{l_{1}}C_{l_{3}}C_{l_{5}}\nonumber\\&+&13\,\,terms, 
\eea 
\bea\fl 
\langle a^{*}_{l_{1}m_{1}}a^{*}_{l_{2}m_{2}}a^{*}_{l_{3}m_{3}}a_{l_{4}m_{4}}\rangle&=&\delta_{l_{1}l_{2}}\delta_{l_{3}l_{4}}\delta_{m_{1}-m_{2}}\delta_{m_{3}m_{4}}(-1)^{m_{1}}C_{l_{1}}C_{l_{3}}\nonumber\\\fl&+&\delta_{l_{1}l_{3}}\delta_{l_{2}l_{4}}\delta_{m_{1}-m_{3}}\delta_{m_{2}m_{4}}(-1)^{m_{1}}C_{l_{1}}C_{l_{2}}\nonumber\\\fl&+&\delta_{l_{3}l_{2}}\delta_{l_{1}l_{4}}\delta_{m_{3}-m_{2}}\delta_{m_{1}m_{4}}(-1)^{m_{3}}C_{l_{1}}C_{l_{3}}, 
\eea 
\bea\fl 
\langle a^{*}_{l_{1}m_{1}} a_{l_{4}m_{4}}\rangle=\delta_{l_{1}l_{4}}\delta_{m_{1}m_{4}}C_{l_{1}}. 
\eea 
The two, four and six point functions respectively provide the following contributions to the Fisher matrix 
\bea\label{one} 
F_{\lambda_{LM}\lambda_{L'M'}}^{(2pf)}=f_{NL}^{2}\left(9\Sigma^{b}_{LM}\right)\delta_{LL'}\delta_{MM'}, 
\eea 
\bea\label{two} 
F_{\lambda_{LM}\lambda_{L'M'}}^{(4pf)}=f_{NL}^{2} \left(-18 \Sigma^{b}_{LM}\right)\delta_{LL'}\delta_{MM'}, 
\eea 
and 
\bea\label{spf} 
F_{\lambda_{LM}\lambda_{L'M'}}^{(6pf)}=f_{NL}^{2}\left(6\Sigma^{a}_{LM}+9\Sigma^{b}_{LM}\right)\delta_{LL'}\delta_{MM'}.   
\eea 
where 
\bea 
\Sigma^{a}_{LM}\equiv \frac{1}{36}\sum_{l_{1},l_{2},l_{3},m_{1},m_{2},m_{3}}\frac{\left(B^{l_{1}l_{2}l_{3}(A)LM}_{m_{1}m_{2}m_{3}}\right)^2}{C_{l_{1}}C_{l_{2}}C_{l_{3}}},\\ 
\Sigma^{b}_{LM}\equiv\frac{1}{36} \sum_{l_{1},l_{2},l_{5},m_{2},m_{5}}\frac{(-1)^{m_{2}+m_{5}}}{C_{l_{1}}C_{l_{2}}C_{l_{5}}}\left(B^{l_{1}l_{2}l_{2}(A)LM}_{0-m_{2}m_{2}}\right)\left(B^{l_{1}l_{5}l_{5}(A)LM}_{0-m_{5}m_{5}}\right) . 
\eea 

The Fisher matrix is given by the sum of Eqs.~(\ref{one}), (\ref{two}) and~(\ref{spf}), and it turns out to be diagonal. Notice that the contributions from $\Sigma^{b}_{LM}$ cancel out.   
The variance of the estimator becomes 
\bea 
\frac{1}{\sigma^2_{\lambda_{LM}}}=\frac{1}{6}\left(3 \mathcal{Q}+6\mathcal{M}\right)_{LM}\, , 
\eea 
where 
\bea\label{fff1}\fl 
\mathcal{Q}&\equiv& \sum_{l_{1}l_{2}l_{3}l_{1}^{'}}\frac{\left(2l_{1}+1\right)\left(2l_{2}+1\right)\left(2l_{3}+1\right)\left(2l_{1}^{'}+1\right)}{(4\pi)^2 C_{l_{1}}C_{l_{2}}C_{l_{3}}}i^{2\left(l_{1}+l_{1}^{'}\right)}\left(b_{l_{1}l_{2}l_{3}}^{l^{'}_{1}}\right)^2\nonumber\\\fl&\times&\left(\begin{array}{ccc}l_{1}^{'}&l_{2}&l_{3}\\0&0&0\end{array}\right)^2\left(\begin{array}{ccc}l_{1}&l_{1}^{'}&L\\0&0&0\end{array}\right)^2\, , 
\eea 
\bea\label{fff2}\fl 
\mathcal{M}&\equiv&\sum_{l_{1}l_{2}l_{3}l_{1}^{'}l_{2}^{''}}\frac{\left(2l_{1}+1\right)\left(2l_{2}+1\right)\left(2l_{3}+1\right)\left(2l_{1}^{'}+1\right)\left(2l_{2}^{''}+1\right)}{(4\pi)^2 C_{l_{1}}C_{l_{2}}C_{l_{3}}}i^{\left(l_{1}+l_{1}^{'}+l_{2}+l_{2}^{''}\right)}b_{l_{1}l_{2}l_{3}}^{l^{'}_{1}}b_{l_{1}l_{2}l_{3}}^{\,\,\,\, l^{''}_{2}}\nonumber\\\fl&\times&\left(\begin{array}{ccc}l_{1}^{'}&l_{2}&l_{3}\\0&0&0\end{array}\right)\left(\begin{array}{ccc}l_{1}&l_{2}^{''}&l_{3}\\0&0&0\end{array}\right)\left(\begin{array}{ccc}l_{1}&l_{1}^{'}&L\\0&0&0\end{array}\right)\left(\begin{array}{ccc}l_{2}&l_{2}^{''}&L\\0&0&0\end{array}\right)\Bigg\{\begin{array}{ccc}l_{1}&l_{1}^{'}&L\\l_{2}&l_{2}^{''}&l_{3}\end{array}\Bigg\}\, , 
\eea 
which are independent from the multipole index $M$. The expression of the bispectra $(b_{l_{1}l_{2}l_{3}}^{l^{'}_{1}})$, 
$( b_{l_{1}l_{2}l_{3}}^{l^{'}_{1}})$, $( b_{l_{1}l_{2}l_{3}}^{\,\,\,\, l^{''}_{2}})$ are given in Eqs.~(\ref{b1}-\ref{b3}). Equations (\ref{10o}) through (\ref{md}) were employed to derive $\mathcal{Q}$ and $\mathcal{M}$. 

We will now evaluate (\ref{fff1}) and (\ref{fff2}) for $L=2$. The triangle inequalities involving $l_{1}$ and $l^{'}_{1}$ are such that $l_{1}^{'}\in[l_{1}-2,l_{1}+2]$; similarly $l_{2}^{''}\in[l_{2}-2,l_{2}+2]$, therefore we expect the approximations $l_{1}^{'} \simeq l_{1}$ and $l_2^{''}\simeq l_{2}$ to be valid in Eqs.~(\ref{fff1}) and (\ref{fff2}) for $l_{1},l_{2}\gg 1$. 

In order to evaluate the sums over $l_{1}$, $l_{2}$ and $l_{3}$, we will assume $l_{1}^{max}\simeq l_{2}^{max}\simeq l_{3}^{max}\simeq l$, which is the maximum multipole to be probed by a given 
experiment. Under these approximations the various bispectra, Eqs.~(\ref{b1}-\ref{b3}), turn out to be all equal and related to the reduced bispectrum $b_{lll}$ given in Eq.~(\ref{Iso}). Using 
Eqs.~(\ref{equal1}) to (\ref{equal2}) listed in Appendix D, we derive 
\bea\label{fq}\fl 
\mathcal{Q}&\simeq& \frac{l^3\left(2 l\right)^4}{(4\pi)^2 C_{l}^{3}}\left(b_{lll}\right)^2\left(\begin{array}{ccc}l_{}&l_{}&l\\0&0&0\end{array}\right)^2\left(\begin{array}{ccc}l_{}&l_{}&2\\0&0&0\end{array}\right)^2\simeq\left(\frac{0.36\times 0.125}{\pi^2}\right)\frac{\left(l^5 b_{lll}\right)^2}{\left(l^2 C_{l}\right)^3}, 
\eea 
\bea\label{fm}\fl 
\mathcal{M}&\simeq& \frac{l^3 \left(2 l\right)^5}{(4\pi)^2 C_{l}^{3}}\left(b_{lll}\right)^2\left(\begin{array}{ccc}l_{}&l_{}&l\\0&0&0\end{array}\right)^2\left(\begin{array}{ccc}l_{}&l_{}&2\\0&0&0\end{array}\right)^2\Bigg\{\begin{array}{ccc}l_{}&l_{}^{}&2\\l_{}&l_{}^{}&l_{}\end{array}\Bigg\}\nonumber\\\fl&\simeq&\left(\frac{2 \times 0.36\times 0.125}{16\pi^2}\right)\frac{\left(l^5 b_{lll}\right)^2}{\left(l^2 C_{l}\right)^3}, 
\eea 
where we replaced the sum over $l_{1}$, $l_{2}$ and $l_{3}$ by a factor $l^3$.   

The final result for the variance of the estimator is 
\bea 
\frac{1}{\sigma_{\lambda_{2M}}} \simeq 0.05 \frac{l^5 b_{lll}}{\left(l^2 C_{l}\right)^{3/2}}. 
\eea 
Using the numerical estimates of $l^2 C_{l}\simeq 6\times 10^{-10}$ and $b_{lll}\simeq (2/3)\times 2.4 \times 10^{-17}  f_{NL} l^{-4}$ from \cite{Komatsu:2001rj}, we finally get 
\bea\label{fr00} 
\frac{1}{\sigma_{\lambda_{2M}}}  \simeq 0.1 f_{NL} \left(\frac{l}{2000}\right). 
\eea   
Notice that in deriving Eqs.~(\ref{fq}) and~(\ref{fm}) under the approximations explained above, the sign of the contribution $Q$ is essentially determined by the sum over $l_1^{'}$ of 
$i^{2 (l_1+l_1^{'})}$. Taking $l_1^{'}\simeq l_1 \simeq l_2 \simeq l_3 \simeq l$, all the other contributions are factored out of this sum and the net result is just a $+1$ contribution as in 
Eq.~(\ref{fq}). A similar argument holds for the sum over $l_1^{'}$ and $l_2^{'}$ in Eq.~(\ref{fff2}). 
By performing numerically the sum over $l_i^{'}$ (keeping all the bispectra equal to $b_{lll}$) and then summing over $l_1,l_2, l_3$, we have checked that the sign and 
the order of magnitude of our results are correct. In fact, in this way, we are able to give an improved computation of the signal-to-noise ratio which does not underestimate the contribution 
from the sums over $l_i^{'}$. We find 
\bea\label{fr01} 
\frac{1}{\sigma_{\lambda_{2M}}}\simeq 0.4 f_{NL} \left(\frac{l}{2000}\right) \, . 
\eea 
For a (full-sky) experiment cosmic variance limited up to $l_{max}=2000$ the secondary effects from lensing are expected to be still subdominant in the $C_l$'s appearing in the denominator of~(\ref{fff1}) and~(\ref{fff2}) \cite{Komatsu:2001rj, BabichZald}. In this case, if we take the central value for the local non-Gaussianity amplitude $f_{NL}=32$ (the constraints from~\cite{komatsuWMAP7} indicate $-10 < f_{NL} < 84$, $95\%$ C.L.), then we see that at $1 \sigma$ the variance for the amplitudes of a quadrupolar statistical anisotropy in the bispectrum $\lambda_{2M}$ turns out to be   $\sigma_{\lambda_{2M}} \simeq 0.08$. If we consider $l_{max}=1500$ we find  $\sigma_{\lambda_{2M}} \simeq 0.1$. This value is roughly representative of an experiment like \emph{Planck} which is signal-dominated up to $l_{max}\simeq 2000$. This means that an experiment like \emph{Planck} could be sensitive 
to an anisotropic bispectrum amplitude of this type as low as $10\%$.

\section{Conclusions}

In this paper we built up an optimal estimator for statistical anisotropy parameters from the CMB bispectrum. We considered models where statistical isotropy is broken in the early Universe by the existence of a preferred spatial direction ($\hat{N}$) due, for instance, to a primordial vector field. The vector field fluctuations can affect, by various mechanisms, the correlation function of curvature fluctuations, by introducing some degree of statistical anisotropy. We parametrized the expression of the anisotropic primordial bispectrum as Eq.~(\ref{bis1}) and derived, from there, the bispectrum of the coefficients of the temperature anisotropies harmonic expansion, Eqs.~(\ref{Iso}), (\ref{5}) and (\ref{6}). In particular, we focused on the simplest possible parametrization of the primordial three-point function, by considering terms of the bispectrum that are proportional to $(\hat{k}\cdot\hat{N})^2$. We also assumed $k$-independent anisotropy parameters. For these parameters, we computed optimal estimators, Eq.~(\ref{qq2}), and their signal-to-noise ratios finding, for a quadrupolar anisotropy, a sensitivity for a 
\emph{Planck}-like experiment to an amplitude of about $10\%$.  
We have pointed out that such an analysis is particularly relevant 
for those models that predict a negligibly small amplitude of the anisotropy in the power spectrum and, at the same time, a potentially high level of statistical anisotropy in the 
bispectrum. In Appendix A we have given some general arguments about this point, and we have considered as an example some details of the vector curvaton model.  But, this can be quite a generic situation (as one can check looking at the results of, e.g., \cite{Yokoyama:2008xw,Dimopoulos:2008yv,Bartolo:2009pa}; see also the recent work~\cite{Shiraishi} on the anisotropic CMB bispectrum in the vector model proposed in~\cite{Yokoyama:2008xw}). For these models analysis focusing only on the power spectrum could result completely blind to such bispectrum anisotropic signatures.

Our analysis is valid for a (full sky) cosmic variance limited experiment up to $l_{\rm max} \simeq 2000$ (without accounting for polarization). We have seen that an experiment like 
\emph{Planck} could be sensitive to a quadrupolar anisotropic bispectrum amplitude as low as $10\%$. Looking forward, we can envisage that the polarization information can help in improving the sensitivity in a similar way as it has been realized for the search of the $f_{\rm NL}$ non-Gaussianity~\cite{BabichZald,MW,Sefusatti,CMBPol,Core}. Of course a realistic search will need to account for various systematic and foregrounds effects (e.g. see Refs.~\cite{Bennett2011,Groeneboom:2009,Hanson:2009gu,Groeneboom:2009cb,Hanson:2010gu,Maris:2010wg}). Also lensing of CMB needs to be taken into account in order to avoid possible anisotropic effects (see, e.g.~\cite{Lewis:2011fk} and Refs. therein). 

Our work could be extended in several directions. For instance, it would be interesting to study the anisotropic corrections expressed in terms of bipolar spherical harmonics (Eq.~(\ref{88})) and their products, so as to generalize our calculation to all known vector field models (thus also including non-Abelian models). Also, a correlation between the bispectrum and the power spectrum statistical anisotropy as well as a k-dependence for the anisotropy parameters could be investigated.

\section*{Acknowledgments} 

NB and ED thank Licia Verde for important discussion and correspondence during the completion of this work. The authors thank the Referee for her/his valuable comments, which resulted in an improved version of this paper. This research has been partially supported by the ASI/INAF Agreement 
I/072/09/0 for the Planck LFI Activity of Phase E2. 

\setcounter{equation}{0} 
\def\theequation{A\arabic{equation}} 
\section*{Appendix A. Primordial vector field models of inflation and the bispectrum.} 
\label{AppendixA}

The bispectrum of curvature fluctuations has been computed in different inflationary models with primordial vector fields; in this paper we  specifically refer to the results from \cite{Yokoyama:2008xw} (a hybrid inflation model with an Abelian vector field coupled to the \textsl{waterfall field}), from \cite{Karciauskas:2008bc} (about a single vector field both in curvaton and hybrid inflation models) and from \cite{Bartolo:2009pa} (on multiple vectors, in particular forming an $SU(2)$ triplet, both within the curvaton model and in standard inflation). A compact expression that well summarizes all of the ``Abelian'' results (in particular 
characterized by only one direction of anisotropy $\hat{N}$) is 
\bea\label{aC}\fl 
B(\vec{k}_{1},\vec{k}_{2},\vec{k}_{3})=B^{iso}(k_{1},k_{2})\Big[1&+&\Gamma \left(\hat{k}_{1}\cdot\hat{N}\right)^2+\Delta\left(\hat{k}_{2}\cdot\hat{N}\right)^2+\Theta\left(\hat{k}_{1}\cdot\hat{N}\right)^2\left(\hat{k}_{2}\cdot\hat{N}\right)^2\nonumber\\\fl&+&\Omega\left(\hat{k}_{1}\cdot\hat{k}_{2}\right)\left(\hat{k}_{1}\cdot\hat{N}\right)\left(\hat{k}_{2}\cdot\hat{N}\right)\Big]+2\,\,perms., 
\eea 
where $B^{iso}$ is a function of the wave vectors lengths ($k_{i} \equiv  |\vec{k}_{i}|$, $i=1,2,3$), that also depends on the parameters of the specific model; the same applies to the functions $\Gamma$, $\Delta$, $\Theta$ and $\Omega$. \\ 
Similarly, the power spectrum has the following form 
\bea\label{ppss} 
P(\vec{k})=P^{iso}(k)\left[1+G(k) \left(\hat{k}\cdot\hat{N}\right)^2\right]. 
\eea 
As mentioned in the introduction, various data analysis have been performed for a power spectrum as in (\ref{ppss}) to find out what the orientation of the preferred direction should be and to set some upper bounds on the amplitude $G$. No such analysis has yet been performed for the bispectrum. \\
Considering a non-Abelian $SU(2)$ model  we can write
an even more general parametrization of the bispectrum (taking for simplicity  three vectors of the same length $N_{A}\equiv |\vec{N}^{a}|$ ($a=1,2,3$))
\bea\label{cr}\fl 
B(\vec{k}_{1},\vec{k}_{2},\vec{k}_{3})=B^{iso}(k_{1},k_{2})\Big[1&+&g_B(k_{1},k_{2})\Big(p(k_{1})\sum_{a}\left(\hat{k}_{1}\cdot\hat{N}^{a}\right)^2+p(k_{2})\sum_{a}\left(\hat{k}_{2}\cdot\hat{N}^{a}\right)^2\nonumber\\\fl&+&p(k_{1})p(k_{2})\hat{k}_{1}\cdot\hat{k}_{2}\sum_{a}\left(\hat{k}_{1}\cdot\hat{N}^{a}\right)\left(\hat{k}_{2}\cdot\hat{N}^{a}\right)\Big)\Big]+2\,\,\,perms.\nonumber\\\fl
&+&g_{c}^{2}H_{*}^{2}\sum_{n}F_{n}(k_{1},k_{2},k_{3})I_{n}(\hat{k}_{i}\cdot\hat{k}_{j},\hat{N}^{a}\cdot\hat{N}^{b},\hat{k}_{i}\cdot\hat{N}^{a}) 
\eea 
where the $SU(2)$ index $a$ runs from $1$ to $3$; the $i,j$ indices in the argument of the $I_{n}$ functions label the three (unit) wave vectors. The isotropic part of the bispectrum is determined by the contributions of both the inflaton and the vector fields, i.e. 
\be 
\label{iso}
B^{iso}(k_{1},k_{2})\equiv N_{\phi}^{2}N_{\phi\phi}P_{\phi}(k_{1})P_{\phi}(k_{2})+N_{A}^{2}N_{AA}P_{+}(k_{1})P_{+}(k_{2})\, ,
\ee
where the power spectra of the fields are weighted by the derivatives of the e-foldings number with respect to the different fields, e.g. $\vec{N}^{a}\equiv(\p N)/(\p \vec{A}^{a})$. Typically $\vec{N}^{a} \propto \vec{A}^{a}$. Also we set $N_{AA}\delta_{ab}\delta_{ij}\equiv(\p^2 N)/(\p A^{a}_{i}\p A^{b}_{j})$.  Here $P_+=(1/2) (P_{\rm R}+P_{\rm L})$ is the mean power spectrum of the two transverse polarization states of the vector field.  Notice that the structure of the first two lines include  the Abelian case (\ref{aC}), whereas the last line of (\ref{cr}) is specific to the non-Abelian case since it includes the contribution from $SU(2)$ interactions ($g_{c}$ is the $SU(2)$ coupling constant). \\
We report the explicit expressions for the functions appearing in (\ref{cr}) 
\bea 
\label{gb}
g_B(k_{1},k_{2})&\equiv& \left(1+\left(\frac{N_{\phi}}{N_{A}}\right)^2\left(\frac{N_{\phi\phi}}{N_{AA}}\right)\frac{P_{\phi}(k_{1})P_{\phi}(k_{2})}{P_{+}(k_{1})P_{+}(k_{2})}\right)^{-1} ,\\ \label{lcr} 
p(k)&\equiv& \frac{P_{l}(k)-P_{+}(k)}{P_{+}(k)} \, ,
\eea 
where $P_l(k)$ is the power spectrum of the longitudinal component.  In the same context, the power spectrum is 
\bea 
\label{ps}
P(\vec{k})=P^{iso}(k)\left[1+\tilde{G}(k)\sum_{a}\left(\hat{k}\cdot\hat{N}^{a}\right)^2\right], 
\eea 
where 
\bea\label{dq} 
\tilde{G}\equiv p(k) g_{P}(k)= p(k) \left(1+\left(\frac{N_{\phi}}{N_{A}}\right)^2\frac{P_{\phi}}{P_{+}}\right)^{-1}. 
\eea 
$P^{iso}(k)\equiv N_{\phi}^{2}P_{\phi}(k)\left[1+\beta (P_{+}/P_{\phi})\right]$ and we introduced the parameter $\beta=\left(N_{A}/N_{\phi}\right)^2$ (the reader can refer to \cite{Bartolo:2009pa} for more details about the quantities introduced in these equations).  Notice the similar 
structure for the power spectrum and bispectrum related quantities, e.g. Eqs.~(\ref{ps})-(\ref{cr}) and Eqs.~(\ref{dq})-(\ref{gb}).

For most vector field models, it is correct to set $p\simeq \mathcal{O}(1)$ and $P_{\phi}\simeq P_{+}$.  Then generally one finds that 
\begin{equation}
\tilde{G} \sim g_P  \sim \frac{P^{({\rm vector})}_\zeta}{P_\zeta} \sim \beta    \, , 
\end{equation}
where $P^{({\rm vector})}_\zeta$ is the (isotropic) contribution of the vector field(s) perturbations to the total (isotropic) curvature perturbation $P_\zeta$ and the last step is valid for $\beta \ll 1$. Similarly for the bispectrum 
\begin{equation}
g_B \sim \frac{B^{(\rm vector)}_\zeta}{B^{(I)}_\zeta}\, ,
\end{equation} 
corresponds to the ratio between the isotropic contribution from the vector field to the total isotropic bispectrum (see Refs.~\cite{Bartolo:2009pa,Bartolo:2009kg} for further details).  No data analysis has been performed for the power spectrum if more than one anisotropy direction is involved; however, we will safely assume that the existing bounds are also valid for a multi-vector case, so $\tilde{G}\ll 1$ and $\beta\ll 1$.\\ 

In the language of the angular decomposition~(\ref{8}), the coefficients $\lambda_{\rm LM}$ turns out to be
\be
\label{stimalambda}
\lambda_{\rm LM} \sim g_B F(\mathbf A) \sim \frac{f_{\rm NL}^{\rm (A)}}{f_{\rm NL}^{(I)}} = 
\frac{f_{\rm NL}^{\rm (A)} }{f_{\rm NL}^{\rm (I) \phi}+f_{\rm NL}^{\rm (I) vector}}\, ,
\ee 
so that they represent the  ratio between the amplitude of the anisotropic (A) bispectrum from the vector fields to the total isotropic (I) bispectrum (which generally receives contribution from both the inflaton and the vector fields).  Here  $F(\mathbf A)$ account for some 
angular configurations which is model-dependent. Just because of this angular configuration dependence, generally it turns out that  $|f_{\rm NL}^{(A)}| \leq |f_{\rm NL}^{(I) \rm vector}|$, so that one expects $ |\lambda_{\rm LM}| \leq 1$.  

\subsection{A worked example}
As an example, consider those models where the contribution to the bispectrum from the inflaton field is negligible (as for the simplest scenarios). In this case 
$f_{\rm NL}^{(I) \phi}$ in Eq.~(\ref{stimalambda}) is negligible and it is easy to obtain from Eq.~(\ref{iso}) 
\begin{equation}
f_{\rm NL}^{(I) \rm vector}= f_{\rm NL}^{(I)} \simeq \frac{N_A^2 N_{AA}}{(N_\phi^2+N_A^2)^2}= \frac{\beta^2}{(1+\beta)^2} \frac{N_{AA}}{N_A^2} \sim \beta^2 
\frac{N_{AA}}{N_A^2} \, . 
\end{equation}   
In the case of the vector curvaton model, for example, one finds that $(N_{AA}/N_A^2) \sim \Omega_A^{-1}$, with $ \Omega_A$ the density parameter of the vector curvaton at the epoch of the curvaton decay, so that 
\be
f_{\rm NL}^{(I)} \simeq \frac{\beta^2}{\Omega_A}\, .
\ee
 According to Eq.~(\ref{stimalambda}) the anisotropic non-Gaussianity amplitude $f_{\rm NL}^{(A)}$ differs from the isotropic one by the 
angular configuration dependence encoded in  $F(\mathbf A)$, which generally is $|F(\mathbf A)| \leq 1$ (this argument agrees with the explicit formulae Eqs.~(34) and (35) of \cite{Dimopoulos:2008yv}). 

At this point we can answer two important questions. First, given that the amplitude of the (isotropic and anisotropic) non-Gaussianity depends on $G^2 \sim \beta^2 \ll 1$, is 
$f_{\rm NL}^{(I)}$ suppressed? This is relevant given our result Eq.~(\ref{fr01}). Second, does a small (non measurable) anisotropy amplitude ($G$) of the power spectrum imply a small (non measurable) amplitude of the anisotropic bispectrum ($\lambda_{\rm LM}$)? 
The answer to both question is actually negative. For example if one takes $G \sim 10^{-2}$ then, in order to have $f_{\rm NL} \sim 30$, we require $\Omega_A \sim 10^{-5}-10^{-6}$, which well satisfies the requirement that in these models the vector field must remain subdominant  ($\Omega_A \ll 1$). Notice that this also satisfies the condition $\Omega_A^2 \geq \beta P_\zeta \sim 10^{-10} \beta$ which comes from the requirement that $|\delta A/ A| \leq 1$ in order to avoid a non-Gaussian perturbation to dominate (see  \cite{Dimopoulos:2008yv}). \footnote{Recall that the vector curvaton contribution to the curvature perturbation is $\zeta_A \simeq \Omega_A (\delta A/A)$ and that $P_\zeta^{iso}=P_\zeta (1+\beta)$.}  This answers also the second question, since the anisotropic non-Gaussianity amplitude 
 $f_{\rm NL}^{(A)}$ differs from the isotropic one by the 
angular configuration dependence  $|F(\mathbf A)| \leq 1$. Therefore it is certainly possible that $f_{\rm NL}^{(A)} \sim f_{\rm NL}^{(I)}$ and therefore to have 
$|\lambda_{\rm LM}| \sim 1$. Our result ~(\ref{fr01}) on the other hand shows that in principle an experiment like Planck can be sensitive to much lower values  of 
$\lambda_{\rm LM}$.

The discussion following Eq.~(\ref{dq}) has been developed with the Abelian contributions to Eq.~(\ref{cr}) in mind. For the non-Abelian contributions the amplitude of the bispectrum is totally independent w.r.t. the power spectrum anisotropy amplitude and the parameter space of the theory is such that a comparable amplitude is possible for $f_{NL}^{(I)}$ and $f_{NL}^{(A)}$ (see \cite{Bartolo:2009pa} for a direct comparison between these contributions). From the expressions of the anisotropy coefficients $I_{n}$ appearing in the non-Abelian contribution \cite{Bartolo:2009pa} it is easy to check that, in order for these contributions to be accounted for in the computation of the CMB bispectrum, one would have to resort to products of bipolar spherical harmonics. 

However, in order to keep our calculations as simple and straighforward as possible in this first paper on statistical anisotropy estimators from the bispectrum, we focused on the Abelian case, considering the Abelian version of the first two lines of (\ref{cr}) (given in Eq.~(\ref{bis1})).

\setcounter{equation}{0} 
\def\theequation{B\arabic{equation}} 
\section*{Appendix B. Fisher matrix/Bayesian analysis for complex parameters.} 

Let us derive the Cramer-Rao inequality for complex parameters following the computation for the real case reviewed, for instance, in Sec.~3.1 of \cite{Heavens:2009nx}. Given two parameters $\lambda_{1}$ and $\lambda_{2}$ and their unbiased estimated values $\hat{\lambda}_{i}$ ($i=1,2$), the following inequality
\bea\label{sin}
\Big\langle\left(\Delta\lambda_{1}+\beta \Delta\lambda_{2}\right)\left(\Delta\lambda_{1}+\beta \Delta\lambda_{2}\right)^{*}\Big\rangle\geq 0,
\eea
(where $\Delta\lambda_{i}\equiv \hat{\lambda}_{i}-\lambda_{i}$) must hold for any value of $\beta$. In particular, for $\beta=-\langle\Delta\lambda_{1}\Delta\lambda_{2}^{*}\rangle/\langle|\Delta\lambda_{2}|^{2}\rangle$, Eq.~(\ref{sin}) provides the \textsl{Schwarz inequality}
\bea
\langle|\Delta\lambda_{1}|^{2}\rangle\langle|\Delta\lambda_{2}|^{2}\rangle\geq |\langle\Delta\lambda_{1}\Delta\lambda_{2}^{*}\rangle|^{2},
\eea
which we can use in the derivation of the Cramer-Rao inequality as follows. Consider for simplicity one parameter only $\lambda$. By definition 
\bea
\langle\Delta\lambda \rangle=\int \Delta\lambda P(x,\lambda) dx=0,
\eea
for some data $x$. Let us derive the previous equation w.r.t. $\lambda$
\bea
\int \left(\Delta\lambda\right)\left(\frac{\p P}{\p \lambda}\right) dx+\int P dx=0,
\eea
then
\bea
\int \left(\Delta\lambda\right) \left(\frac{\p \ln P}{\p\lambda}\right) P dx=\langle \left(\Delta\lambda\right) \left(\frac{\p\ln P}{\p\lambda}\right)\rangle=-1,
\eea
where we used $\p P/\p \lambda=P\left(\p \ln P/\p \lambda\right)$ and $\int P dx=1$. From the Schwarz inequality 
\bea
\Big\langle\left|\left(\Delta\lambda\right) \left(\frac{\p\ln P}{\p\lambda}\right)\right|^{2}\Big\rangle\leq \Big\langle\left|\Delta\lambda\right|^{2}\Big\rangle \Big\langle\left|\frac{\p\ln P}{\p\lambda}\right|^{2}\Big\rangle,\nonumber
\eea
therefore
\bea\label{s.heav}
\Big\langle\left|\Delta\lambda\right|^{2}\Big\rangle\geq \frac{1}{\Big\langle\left|\frac{\p\ln P}{\p\lambda}\right|^{2}\Big\rangle}.
\eea
From
\bea\fl
0=\frac{\p}{\p\lambda}\frac{\p}{\p \lambda^{*}}\int P dx=\int  \left(\frac{\p\ln  P}{\p\lambda^{}}\right)\left(\frac{\p \ln P}{\p\lambda^{*}}\right)P dx+\int  \left(\frac{\p^{2} \ln P}{\p\lambda^{}\p\lambda^{*}}\right)P dx,
\eea
we get
\bea\fl
\Big\langle  \left(\frac{\p \ln P}{\p\lambda^{}}\right)\left(\frac{\p \ln P}{\p\lambda^{*}}\right) \Big\rangle=-\Big\langle \left(\frac{\p^{2} \ln P}{\p\lambda^{}\p\lambda^{*}}\right)  \Big\rangle
\eea
which, after replacement in Eq.~(\ref{s.heav}), gives
\bea
\Big\langle\left|\Delta\lambda\right|^{2}\Big\rangle\geq -\frac{1}{\Big\langle\frac{\p^2\ln P}{\p\lambda\p\lambda^{*}}\Big\rangle}=\frac{1}{F},
\eea
where $F$ is the Fischer matrix for $\lambda$. 

\setcounter{equation}{0} 
\def\theequation{C\arabic{equation}} 
\section*{Appendix C. Higher order terms in the Edgeworth expansion.} 

The Edgeworth expansion is given by
\bea\label{toadd}\fl
P(a)&=&\Big(1-\frac{1}{3!}\sum_{l_{i},m_{i}}C^{l_{1}l_{2}l_{3}}_{m_{1}m_{2}m_{3}}\frac{\p}{\p a_{l_{1}m_{1}}}\frac{\p}{\p a_{l_{2}m_{2}}}\frac{\p}{\p a_{l_{3}m_{3}}}\nonumber\\\fl&+&\frac{1}{4!}\sum_{l_{i},m_{i}}C^{l_{1}l_{2}l_{3}l_{4}}_{m_{1}m_{2}m_{3}m_{4}}\frac{\p}{\p a_{l_{1}m_{1}}}\frac{\p}{\p a_{l_{2}m_{2}}}\frac{\p}{\p a_{l_{3}m_{3}}}\frac{\p}{\p a_{l_{4}m_{4}}}\nonumber\\\fl&-&\frac{1}{5!}\sum_{l_{i},m_{i}}C^{l_{1}l_{2}l_{3}l_{4}l_{5}}_{m_{1}m_{2}m_{3}m_{4}m_{5}}\frac{\p}{\p a_{l_{1}m_{1}}}\frac{\p}{\p a_{l_{2}m_{2}}}\frac{\p}{\p a_{l_{3}m_{3}}}\frac{\p}{\p a_{l_{4}m_{4}}}\frac{\p}{\p a_{l_{5}m_{5}}}\nonumber\\\fl&+&\frac{1}{6!}\sum_{l_{i},m_{i}}C^{l_{1}l_{2}l_{3}l_{4}l_{5}l_{6}}_{m_{1}m_{2}m_{3}m_{4}m_{5}m_{6}}\frac{\p}{\p a_{l_{1}m_{1}}}\frac{\p}{\p a_{l_{2}m_{2}}}\frac{\p}{\p a_{l_{3}m_{3}}}\frac{\p}{\p a_{l_{4}m_{4}}}\frac{\p}{\p a_{l_{5}m_{5}}}\frac{\p}{\p a_{l_{6}m_{6}}}+...\Big)\nonumber\\\fl&\times&\frac{e^{-\frac{1}{2}\sum_{l_{q},m_{q}}a^{*}_{l_{4}m_{4}}C^{-1}_{l_{4}m_{4},l_{5}m_{5}}a_{l_{5}m_{5}}}}{\left(2\pi\right)^{N_{p}/2}\left(detC\right)^{1/2}},
\eea
where $q=4,5$ and
\bea\fl
C^{l_{1}l_{2}l_{3}}_{m_{1}m_{2}m_{3}}&\equiv&\langle a_{l_{1}m_{1}}a_{l_{2}m_{2}}a_{l_{3}m_{3}} \rangle ,\\\fl
C^{l_{1}l_{2}l_{3}l_{4}}_{m_{1}m_{2}m_{3}m_{4}}&\equiv& \langle  a_{l_{1}m_{1}}a_{l_{2}m_{2}}a_{l_{3}m_{3}}a_{l_{4}m_{4}} \rangle_{c} , \\\fl
C^{l_{1}l_{2}l_{3}l_{4}l_{5}}_{m_{1}m_{2}m_{3}m_{4}m_{5}}&\equiv& \langle a_{l_{1}m_{1}}a_{l_{2}m_{2}}a_{l_{3}m_{3}}a_{l_{4}m_{4}}a_{l_{5}m_{5}} \rangle_{c} , \\\label{sixpf}\fl
C^{l_{1}l_{2}l_{3}l_{4}l_{5}l_{6}}_{m_{1}m_{2}m_{3}m_{4}m_{5}m_{6}}&\equiv&\langle a_{l_{1}m_{1}}a_{l_{2}m_{2}}a_{l_{3}m_{3}}a_{l_{4}m_{4}}a_{l_{5}m_{5}}a_{l_{6}m_{6}} \rangle_{c} \\\fl&+&\Big[\langle a_{l_{1}m_{1}}a_{l_{2}m_{2}}a_{l_{3}m_{3}}\rangle\langle a_{l_{4}m_{4}}a_{l_{5}m_{5}}a_{l_{6}m_{6}}\rangle+9\,\,perms.\Big]\nonumber\\\fl
C^{l_{1}l_{2}l_{3}l_{4}l_{5}l_{6}l_{7}}_{m_{1}m_{2}m_{3}m_{4}m_{5}m_{6}m_{7}}&\equiv&\langle a_{l_{1}m_{1}}a_{l_{2}m_{2}}a_{l_{3}m_{3}}a_{l_{4}m_{4}}a_{l_{5}m_{5}}a_{l_{6}m_{6}}a_{l_{7}m_{7}} \rangle_{c} \\\fl&+& \Big[\langle a_{l_{1}m_{1}}a_{l_{2}m_{2}}a_{l_{3}m_{3}}a_{l_{4}m_{4}}\rangle_{c}\langle a_{l_{5}m_{5}}a_{l_{6}m_{6}}a_{l_{7}m_{7}}\rangle+34\,\,perms.\Big]\nonumber
\eea
and so on for the rest of the series (the subscripts $c$ in the previous equations label the fully connected correlators).\\
We want to perform our computation of the estimator and of the Fisher matrix for the $\lambda_{LM}$s coefficients to leading order in powers of $f_{NL}$; this implies that we can immediately drop all the terms in the Edgeworth expansion with coefficients that are proportional to $(f_{NL})^{k}$, $k\geq 3$ (see e.g. Eqs.~(\ref{qq1})-(\ref{qq2})). Moreover, we choose to neglect the four-point function and all higher order (connected) correlators. With these premises, the only terms left in the Edgeworth expansion are the first line of Eq.~(\ref{toadd}) and the second line of Eq.~(\ref{sixpf})
\bea\label{toadd3}\fl
P(a)&=&\Big(1-\frac{1}{3!}\sum_{l_{i},m_{i}}\langle a_{l_{1}m_{1}}a_{l_{2}m_{2}}a_{l_{3}m_{3}} \rangle\frac{\p}{\p a_{l_{1}m_{1}}}\frac{\p}{\p a_{l_{2}m_{2}}}\frac{\p}{\p a_{l_{3}m_{3}}}\nonumber\\\fl&+&\frac{1}{72}\sum_{l_{i},m_{i}}\langle a_{l_{1}m_{1}}a_{l_{2}m_{2}}a_{l_{3}m_{3}}\rangle\langle a_{l_{4}m_{4}}a_{l_{5}m_{5}}a_{l_{6}m_{6}}\rangle\frac{\p}{\p a_{l_{1}m_{1}}}\frac{\p}{\p a_{l_{2}m_{2}}}\frac{\p}{\p a_{l_{3}m_{3}}}\frac{\p}{\p a_{l_{4}m_{4}}}\frac{\p}{\p a_{l_{5}m_{5}}}\frac{\p}{\p a_{l_{6}m_{6}}}\Big)\nonumber\\\fl&\times&\frac{e^{-\frac{1}{2}\sum_{l_{q},m_{q}}a^{*}_{l_{4}m_{4}}C^{-1}_{l_{4}m_{4},l_{5}m_{5}}a_{l_{5}m_{5}}}}{\left(2\pi\right)^{N_{p}/2}\left(detC\right)^{1/2}}.
\eea
The next step is to compute the derivatives of the Gaussian PDF; for a diagonal variance we find
\bea\label{toadd4}\fl
D^{l_{1}l_{2}l_{3}}_{m_{1}m_{2}m_{3}}&\equiv&-\left[\frac{e^{-\frac{1}{2}\sum_{l_{},m_{}}a^{*}_{l_{}m_{}}C^{-1}_{l_{}}a_{l_{}m_{}}}}{\left(2\pi\right)^{N_{p}/2}\left(detC\right)^{1/2}}\right]^{-1} \frac{\p}{\p a_{l_{1}m_{1}}}\frac{\p}{\p a_{l_{2}m_{2}}}\frac{\p}{\p a_{l_{3}m_{3}}}\left(e^{-\frac{1}{2}\sum_{l_{},m_{}}a^{*}_{l_{}m_{}}C^{-1}_{l_{}}a_{l_{}m_{}}}\right)\nonumber\\\fl&=&\frac{a^{*}_{l_{1}m_{1}}a^{*}_{l_{2}m_{2}}a^{*}_{l_{3}m_{3}}}{C_{l_{1}}C_{l_{2}}C_{l_{3}}}-\frac{(-1)^{m_{2}}}{C_{l_{1}}C_{l_{2}}}\delta_{l_{2}l_{3}}\delta_{m_{2}-m_{3}}a^{*}_{l_{1}m_{1}}\nonumber\\\fl&-&\frac{(-1)^{m_{1}}}{C_{l_{1}}C_{l_{2}}}\delta_{l_{1}l_{3}}\delta_{m_{1}-m_{3}}a^{*}_{l_{2}m_{2}}-\frac{(-1)^{m_{1}}}{C_{l_{1}}C_{l_{3}}}\delta_{l_{1}l_{2}}\delta_{m_{1}-m_{2}}a^{*}_{l_{3}m_{3}},\\\fl
D^{l_{1}l_{2}l_{3}l_{1}l_{2}l_{3}l_{4}l_{5}l_{6}}_{m_{1}m_{2}m_{3}m_{4}m_{5}m_{6}}&\equiv&\left[\frac{e^{-\frac{1}{2}\sum_{l_{},m_{}}a^{*}_{l_{}m_{}}C^{-1}_{l_{}}a_{l_{}m_{}}}}{\left(2\pi\right)^{N_{p}/2}\left(detC\right)^{1/2}}\right]^{-1}  \frac{\p}{\p a_{l_{1}m_{1}}}\frac{\p}{\p a_{l_{2}m_{2}}}\frac{\p}{\p a_{l_{3}m_{3}}}\frac{\p}{\p a_{l_{4}m_{4}}}\frac{\p}{\p a_{l_{5}m_{5}}}\nonumber\\\fl&\times&\frac{\p}{\p a_{l_{6}m_{6}}}\left(e^{-\frac{1}{2}\sum_{l_{},m_{}}a^{*}_{l_{}m_{}}C^{-1}_{l_{}}a_{l_{}m_{}}}\right)= \frac{a^{*}_{l_{1}m_{1}}a^{*}_{l_{2}m_{2}}a^{*}_{l_{3}m_{3}}a^{*}_{l_{4}m_{4}}a^{*}_{l_{5}m_{5}}a^{*}_{l_{6}m_{6}}}{C_{l_{1}}C_{l_{2}}C_{l_{3}}C_{l_{4}}C_{l_{5}}C_{l_{6}}}\nonumber\\\fl&-&\Big(\frac{a^{*}_{l_{1}m_{1}}a^{*}_{l_{2}m_{2}}a^{*}_{l_{3}m_{3}}a^{*}_{l_{4}m_{4}}}{C_{l_{1}}C_{l_{2}}C_{l_{3}}C_{l_{4}}C_{l_{5}}}(-1)^{m_{5}}\delta_{l_{5}l_{6}}\delta_{m_{5}-m_{6}}+14\,\,perms.\Big)\nonumber\\\fl&+&\Big(\frac{a^{*}_{l_{1}m_{1}}a^{*}_{l_{6}m_{6}}}{C_{l_{1}}C_{l_{2}}C_{l_{3}}C_{l_{6}}}(-1)^{m_{2}+m_{3}}\delta_{l_{2}l_{5}}\delta_{m_{2}-m_{5}}\delta_{l_{3}l_{4}}\delta_{m_{3}-m_{4}}+44\,\,perms.\Big)\nonumber\\\fl&-&\Big(\frac{(-1)^{m_{1}+m_{2}+m_{3}}}{C_{l_{1}}C_{l_{2}}C_{l_{3}}}\delta_{l_{2}l_{5}}\delta_{m_{2}-m_{5}}\delta_{l_{3}l_{4}}\delta_{m_{3}-m_{4}}\delta_{l_{1}l_{6}}\delta_{m_{1}-m_{6}}+14\,\,perms.\Big).
\eea
For small non-Gaussianity, Eq.~(\ref{pdfexp}) then reads
\bea\label{toadd2} 
\ln(P)&=&\ln(1+x+y)-\frac{1}{2}\sum_{l,m}a^{*}_{l_{}m_{}}C^{-1}_{l}a_{l_{}m_{}}+const.\nonumber\\&\simeq& x+y-\frac{x^2}{2}-\frac{1}{2}\sum_{l,m}a^{*}_{l_{}m_{}}C^{-1}_{l}a_{l_{}m_{}}+const., 
\eea 
where 
\bea\label{ydef}\fl
x&\equiv& \frac{f_{NL}}{6}\sum_{l_{i}m_{i}}\left[B_{m_{1}m_{2}m_{3}}^{l_{1}l_{2}l_{3}(I)}|_{f_{NL}=1}+\sum_{LM}\lambda_{LM} B_{m_{1}m_{2}m_{3}}^{l_{1}l_{2}l_{3}(LM)}|_{f_{NL}=1}\right]D^{l_{1}l_{2}l_{3}}_{m_{1}m_{2}m_{3}},\nonumber\\\fl y&\equiv& \frac{f_{NL}^{2}}{72}\sum_{l_{i}m_{i}}\left[B_{m_{1}m_{2}m_{3}}^{l_{1}l_{2}l_{3}(I)}|_{f_{NL}=1}+\sum_{LM}\lambda_{LM} B_{m_{1}m_{2}m_{3}}^{l_{1}l_{2}l_{3}(LM)}|_{f_{NL}=1}\right]\nonumber\\\fl &\times&\left[B_{m_{4}m_{5}m_{6}}^{l_{4}l_{5}l_{6}(I)}|_{f_{NL}=1}+\sum_{\tilde{L}\tilde{M}}\lambda_{\tilde{L}\tilde{M}} B_{m_{4}m_{5}m_{6}}^{l_{4}l_{5}l_{6}(\tilde{L}\tilde{M})}|_{f_{NL}=1}\right]D^{l_{1}l_{2}l_{3}l_{4}l_{5}l_{6}}_{m_{1}m_{2}m_{3}m_{4}m_{5}m_{6}}
\eea
($x$ was also defined in Eq.~(\ref{xdef})).
Let us derive Eq.~(\ref{toadd2}) w.r.t. $\lambda_{LM}$
\begin{equation}\label{derP}
\frac{\partial \ln P}{\partial \lambda_{LM}}=\frac{\partial}{\partial \lambda_{LM}}\left(x-\frac{x^2}{2}\right)+\frac{\partial}{\partial \lambda_{LM}}\left(y\right). 
\end{equation} 
The derivative of the first bracket on the r.h.s of the previous equation is given by the r.h.s. of Eq.~(\ref{derp}); the derivative of the second bracket is
\bea\fl
\frac{\partial}{\partial \lambda_{LM}}\left(y\right)&=&\frac{f_{NL}^{2}}{36}\sum_{l_{i}m_{i}}\Big[B_{m_{1}m_{2}m_{3}}^{l_{1}l_{2}l_{3}(I)}|_{f_{NL}=1}B_{m_{4}m_{5}m_{6}}^{l_{4}l_{5}l_{6}(LM)}|_{f_{NL}=1}\nonumber\\\fl&+&\sum_{\tilde{L}\tilde{M}}\lambda_{\tilde{L}\tilde{M}} B_{m_{1}m_{2}m_{3}}^{l_{1}l_{2}l_{3}(LM)}|_{f_{NL}=1}B_{m_{4}m_{5}m_{6}}^{l_{4}l_{5}l_{6}(\tilde{L}\tilde{M})}|_{f_{NL}=1}\Big]D^{l_{1}l_{2}l_{3}l_{4}l_{5}l_{6}}_{m_{1}m_{2}m_{3}m_{4}m_{5}m_{6}}.
\eea
We can easily show that the $y$ term in Eq.~(\ref{toadd2}) does not affect the estimators of the $\lambda_{LM}$ coefficients: Eq.~(\ref{derp}) becomes
\bea\fl
\frac{\partial \ln P}{\partial \lambda_{LM}}=f_{NL} B^{(A)LM} -f_{NL}^2 B^{(I)} B^{(A)LM}-f_{NL}^2 \lambda^*_{L'M'} {B^{(A)L'M' }}^* B^{(A)LM}+\frac{\partial}{\partial \lambda_{LM}}\left(y\right);
\eea
the contribution from the derivative of $y$ is null once one replaces it with its expectation value, exactly like it was done for instance for the term ${B^{(A)L'M'}}^* B^{(A)LM}$ (see comment after Eq.~(\ref{derp})). In fact, it is easy to check that $\langle D^{l_{1}l_{2}l_{3}l_{4}l_{5}l_{6}}_{m_{1}m_{2}m_{3}m_{4}m_{5}m_{6}}\rangle=0$ at zeroth order in $f_{NL}$ and neglecting the connected four and six-point functions.\\
The contribution of the $y$ term to the Fisher matrix is
\bea\fl
F_{\lambda_{LM}\lambda_{L'M'}}&\supset&-\Bigg\langle\frac{\p^2 y}{\p\lambda^{*}_{L'M'}\p \lambda_{LM}} \Bigg\rangle\\\fl&=&-\frac{f_{NL}^{2}}{36}\sum_{l_{i}m_{i}}\Big[(-1)^{M'}B_{m_{1}m_{2}m_{3}}^{l_{1}l_{2}l_{3}(LM)}|_{f_{NL}=1}B_{m_{4}m_{5}m_{6}}^{l_{4}l_{5}l_{6}(L'-M')}|_{f_{NL}=1}\Big]\langle D^{l_{1}l_{2}l_{3}l_{4}l_{5}l_{6}}_{m_{1}m_{2}m_{3}m_{4}m_{5}m_{6}}\rangle,\nonumber
\eea
therefore the Fisher matrix has no leading order contributions other than the ones already derived in Sec.~(3).\\

We have thus shown that, if one neglects all connected correlators beyond third order and wants to keep the leading order contributions in $f_{NL}$ to the Fisher matrix and to the estimator, the Edgeworth expansion can be truncated as in Sec.~(3).

\setcounter{equation}{0} 
\def\theequation{D\arabic{equation}} 
\section*{Appendix D. Useful relations involving the Wigner 3-J symbols.} 

We will list the identities involving the Wigner 3-J symbol that we used in our calculations 

\begin{equation}\label{10o}\fl 
\sum_{m} 
(-1)^{l+m} 
\left(\begin{array}{ccc}l&l&L\\m&-m&0\end{array}\right) 
= 
\delta_{L0}\sqrt{\frac{2l+1}{2L+1}}, 
\end{equation} 

\begin{equation}\label{11o}\fl 
\left(\begin{array}{ccc}l&l&0\\m&-m&0\end{array}\right) 
= \frac{(-1)^{l-m}}{\sqrt{2l+1}}. 
\end{equation} 

\bea\fl 
\sum_{m_{i}}\left(\begin{array}{ccc}l_{1}^{}&l_{2}&l_{3}\\m_{1}&m_{2}&m_{3}\end{array}\right)^2=1, 
\eea 
\bea\fl 
\sum_{m_{2},m_{3}}\left(\begin{array}{ccc}l_{1}^{'}&l_{2}&l_{3}\\m_{1}^{'}&m_{2}&m_{3}\end{array}\right)\left(\begin{array}{ccc}l_{1}^{''}&l_{2}&l_{3}\\m_{1}^{''}&m_{2}&m_{3}\end{array}\right)=\frac{\delta_{l_{1}^{'}l_{1}^{''}}\delta_{m_{1}^{'}m_{1}^{''}}}{(2l^{'}_{1}+1)}, 
\eea 
\bea\label{md}\fl 
\sum_{m_{1},m_{2},m_{4},m_{5},m_{6}}(-1)^{l_{1}+l_{2}+l_{4}+l_{5}+l_{6}-m_{1}-m_{2}-m_{4}-m_{5}-m_{6}}\left(\begin{array}{ccc}l_{2}&l_{3}&l_{1}\\m_{2}&-m_{3}&m_{1}\end{array}\right)\left(\begin{array}{ccc}l_{1}&l_{5}&l_{6}\\-m_{1}&m_{5}&m_{6}\end{array}\right)\nonumber\\\fl\times\left(\begin{array}{ccc}l_{5}&l_{3}^{'}&l_{4}\\-m_{5}&m_{3}^{'}&m_{4}\end{array}\right)\left(\begin{array}{ccc}l_{4}&l_{2}&l_{6}\\-m_{4}&-m_{2}&-m_{6}\end{array}\right)=\frac{(-1)^{l_{3}-m_{3}}}{2l_{3}+1}\delta_{l_{3}l_{3}^{'}}\delta_{m_{3}m_{3}^{'}}\Bigg\{\begin{array}{ccc}l_{1}&l_{2}^{}&l_{3}\\l_{4}&l_{5}^{}&l_{6}\end{array}\Bigg\} 
\eea 
\bea\label{equal1}\fl 
l^3\left(\begin{array}{ccc}l_{}&l_{}&l\\0&0&0\end{array}\right)^2\simeq 0.36 l \,\,\,\,\,(l\gg 1;\,\, l\,\,even)
\eea 
\bea\label{equal3}\fl 
\left(\begin{array}{ccc}l_{}&l_{}&2\\0&0&0\end{array}\right)^2\simeq \frac{0.125}{l}\,\,\,\,\,\,\,\,(l\gg 1) 
\eea 
\bea\label{equal2}\fl 
\Bigg|\Bigg\{\begin{array}{ccc}l_{}&l&2\\l&l&l\end{array}\Bigg\}\Bigg|\simeq \frac{1}{16 l} \,\,\,\,\,\,\,\,\,\,\,\,\,\,\,(l\gg 1) 
\eea

\end{document}